\documentclass[reprint]{JASA}

\usepackage{pifont}

\newcommand{\xmark}{\ding{55}}%

\begin{document}

\title[]{Advanced accent/dialect identification and accentedness assessment with multi-embedding models and automatic speech recognition}
\author{Shahram Ghorbani}
\author{~John~H.L. Hansen}
\affiliation{Center for Robust Speech Systems (CRSS), University of Texas at Dallas, Richardson, TX 75080, USA}

%

\date{\today} 

\begin{abstract}


Accurately classifying accents and assessing accentedness in non-native speakers are both challenging tasks due to the complexity and diversity of accent and dialect variations. In this study, embeddings from advanced pre-trained language identification (LID) and speaker identification (SID) models are leveraged to improve the accuracy of accent classification and non-native accentedness assessment. Findings demonstrate that employing pre-trained LID and SID models effectively encodes accent/dialect information in speech. Furthermore, the LID and SID encoded accent information complement an end-to-end accent identification (AID) model trained from scratch. By incorporating all three embeddings, the proposed multi-embedding AID system achieves superior accuracy in accent identification. Next, we investigate leveraging automatic speech recognition (ASR) and accent identification models to explore accentedness estimation. The ASR model is an end-to-end connectionist temporal classification (CTC) model trained exclusively with en-US utterances. The ASR error rate and en-US output of the AID model are leveraged as objective accentedness scores. Evaluation results demonstrate a strong correlation between the scores estimated by the two models. Additionally,  a robust correlation between the objective accentedness scores and subjective scores based on human perception is demonstrated,  providing evidence for the reliability and validity of utilizing AID-based and ASR-based systems for accentedness assessment in non-native speech.

\end{abstract}


\maketitle

\section{\label{sec:1} Introduction}

The ability to accurately classify and assess accentedness of non-native speech is a crucial aspect for speech technology and natural language processing. Accent classification refers to identifying the accent or dialect of a speaker based on their speech, and accentedness assessment refers to measuring the degree of accent present in a speaker's voice. Both accent classification and accentedness assessment have essential applications in areas such as speech recognition, dialog systems, speech synthesis, and second language acquisition.

The accent of non-native speakers has a significant impact on recognition performance for automatic speech recognition (ASR) systems \citep{domainexpansion}. To improve ASR for accented speech, accent classification can be used as a preceding step, providing accent information that can be leveraged during training, model selection, or simply decoding \citep{viglino2019end}. This is particularly important in multilingual multi-accent environments, where the system needs to recognize speech from speakers with diverse accents. Additionally, accent classification can be used to improve speech synthesis systems by generating speech that more closely mimics the accent of the desired target speaker.

The assessment of accentedness has crucial applications in the field of second language acquisition. It provides valuable insights into the degree of accent present in a non-native speaker's voice, which is particularly important given the strong correlation between level of accentedness and intelligibility of second language (L2) speech \citep{van2016native}. For second language learners, accurately measuring and tracking their progress in reducing accent can be an important motivator to help learners focus on adjusting specific areas of their pronunciation. Additionally, measuring accentedness information assists language teachers and educators in identifying and addressing specific accent-related challenges for their students.



Several studies have investigated the relationship between accent and variations in speech elements such as spectral features (e.g., formant frequencies) and temporal features (e.g., intonation and duration) \citep{arslan1997frequency,ferragne2010formant,arslan1996language}. These studies have eventually led to the development of various statistical models and machine-learning techniques to automatically classify accents based on these variations. For example, in \citep{AID2005},  second and third formant frequencies were used to differentiate between American English and Indian English accents, achieving successful classification results with a Gaussian mixture model (GMM) classification method. In a similar study \citep{ghesquiere2002flemish}, a text-dependent model utilized formant and phoneme duration features to distinguish between accents of Antwerp and Brabant and accents of other Dutch provinces. Piat et al.  \citep{piat2008foreign} proposed a statistical method utilizing prosodic factors and demonstrated that duration and energy of speech are effective features for accent classification.

Recently, there have been developments in the field of accent classification utilizing deep neural network-based approaches. One example is leveraging bi-directional Long Short-Term Memory (bLSTM) networks to model the longer-term acoustic context \citep{weninger2019deep}.  In \citep{jiao2016accent}, a combination of long-term and short-term training for accent classification was proposed where deep neural networks (DNNs) were trained on long-term statistical features, and recurrent neural networks (RNNs) trained on short-term acoustic features. In a recent study, \citep{huang2021aispeech}, the efficacy of the phone posteriorgram (PPG) feature for accent identification was investigated. Additionally, they leveraged text-to-speech (TTS) technology to augment limited accented data to generate new accented model training data.


The study of accents in non-native speakers of a language (L2) involves analyzing the influence of their native language (L1) on phonetics, prosody, pronunciation, and speech patterns in the non-primary foreign language.  Traditionally, assessment of accents in non-native speakers relies on native listeners who judge accentedness of speech samples. The objective of automatic accent evaluation is to develop algorithms that can automatically identify the key features that contribute to a speaker's accent. Understanding the contributing factors to accented speech has resulted in techniques to detect indicators of accented speech automatically. Several previous studies have been conducted to determine characteristics in accented speech, such as segmental, prosodic, and fluency features, that impact speech intelligibility \citep{kang2020}. Studies have also isolated phonetic and phonological aspects as primary factors affecting perception of accent \citep{porretta2012predicting}. Alternatively, suprasegmental features such as fundamental frequency range and word emphasis have also been demonstrated to have a separate impact on accent perception \citep{kang2010relative}.

 In recent years, ASR systems have been utilized to evaluate pronunciation accuracy at the segmental level. Such ASR systems have been applied in developing automated systems, referred to as mispronunciation detection and diagnosis (MDD). MDDs can identify mispronunciations made by language learners and provide student based diagnostic feedback. To achieve MDD systems, Log-Likelihood Ratio (LLR) was utilized in previous research to distinguish between native-like and non-native speech \citep{franco1999automatic}. Witt and Young \citep{witt2000phone} proposed the "Goodness of Pronunciation" (GOP), which is a log-likelihood ratio score that considers the probability of both the intended canonical phone and alternative phones. 
Recent advancements in MDD systems have also adopted state-of-the-art ASR systems for phone recognition trained on supervised and unsupervised data. These advanced systems have improved the detection of deviations from the expected phonetic production for specific accents~\citep{yang2022improving,wu2021transformer}.

One key challenge in accent classification and accentedness assessment is the variability and complexity of accent. There exists a wide range of accent and dialect variations across different L1-L2 language pairs. This study has employed American English (en-US) as the assumed target language. Even within the same language, there can also be significant variations based on regional, social, and individual factors. This variability makes it difficult to develop comprehensive models that can accurately classify and assess accentedness. Additionally, the availability of training data for accented speech is limited due to several factors, including the lack of standardized accent annotation in existing speech corpora, and the cost and difficulty of collecting and annotating accented speech data. 

This study proposes innovative solutions for accent/dialect classification and assessment that also mitigate the issue of limited available training data. While training accented data is limited, related tasks such as language identification (LID) and speaker identification (SID) leverage large amounts of data to train advanced models. For example, for the LID task, VoxLingua107 is a corpus of data from 107 languages with an average per language duration of 62 hours~\citep{valk2021slt}, while VoxCeleb contains over a million utterances from over 7k speakers for speaker identification~\citep{voxceleb2}. In light of a recent study \citep{LIDaccent}, which found that language identification (LID)  performance significantly drops when evaluating speech from non-native or regional accents, here, we investigate LID embedding and demonstrate that it can encode distinct representations for native and non-native speakers of a language. Moreover, SID embeddings also capture speaker characteristics, such as accent, gender, stress, emotion, style, age, nationality, and language \citep{probing,disentangled}. Therefore, in this study, we explore the potential of using embeddings from advanced LID and SID systems as auxiliary features to train improved accent/dialect identification systems. By leveraging these pre-trained models, we can mitigate the need for large amounts of accented training data, thus achieving trained accent classification models with greater accuracy.

 To compare performance of accent classification models, we study their performance and compare with solutions based on an End-to-End accent classification model trained from scratch. Our findings will demonstrate that accent embeddings from LID and SID models encode complementary information compared to an E2E accent identification (AID) model. Based on these results, we propose a multi-embedding AID (ME-AID) system that leverages these three embeddings and demonstrates superior accent classification accuracy.


For accentedness estimation, we investigate leveraging two automatic assessment systems; AID-based and ASR-based. Specifically, the AID model is trained to classify en-US utterances from non-native ones, and the en-US score is used as an objective measure. As such, the higher the score, the lower the estimated level of accentedness. For ASR-based system accentedness estimation, we use an ASR error rate as an accentedness metric. The objective is to establish a high correlation between recognition error-rate and accentedness level. To ensure that ASR error-rate accurately reflects the accentedness of the input speech, we suggest training an ASR model exclusively with en-US transcribed speech. Another aspect of the ASR model that contributes to this objective is not to model semantic context, but instead rely solely on the input signal. Among E2E ASR systems, we leverage a connectionist temporal classification (CTC)-based model, since CTC does not model the inter-label dependencies~\citep{CTC}. Therefore, an E2E CTC-based model is trained with speech data from native en-US speakers.

Furthermore, we examine the correlation between accentedness scores generated by ASR and AID systems, and subjective scores based on human listener perception. The computed Pearson coefficient between the AID-based score and the subjective score is 0.68, while the correlation between the ASR-based score and the subjective score here is -0.78, indicating a strong correlation between the objective and subjective scores. These findings provide compelling evidence for the reliability and validity of utilizing AID-based and ASR-based scores as an overall objective measure of accentedness for non-native speech.

 Key contributions of this study are summarized as follows:
\begin{enumerate}
\itemsep0em 
  \item A novel multi-embedding approach for accent/dialect classification by leveraging pre-trained LID and SID models to improve accuracy and address the challenge of limited accented speech training data.
\item Leveraging the multi-embedding AID model for accentedness estimation for non-native en-US speech. Investigation of the correlation between estimated scores and human-perceived scores, revealing a strong association between the two.
\item Introduction of an End-to-End ASR system for estimating accentedness in speech. Results demonstrate a strong correlation between objective scores produced by the model and subjective scores perceived by humans.
\item  A systematic investigation of the consistency between accentedness scores estimated by AID and ASR systems, providing evidence for the reliability and validity of these objective measures for accentedness in non-native speech.
\end{enumerate}

The remainder of this paper is organized as follows: Sec.~\ref{sec:2} proposes the accent classification approaches. Sec.~\ref{sec:3} considers our objective accentedness estimating approaches. Sec.~\ref{sec:4} presents the experimental setup and results, and  Sec.~\ref{sec:5} dicusses the impact and draws conclusions from this study.

\section{\label{sec:2} Accent/Dialect Classification Systems}

This section describes accent/dialect classification approaches investigated. First, an overview is presented of the advanced LID and SID models utilized to generate accent embeddings, followed by the proposed end-to-end and multi-embedding accent/dialect classification methods.

\subsection{\label{subsec:2:3} End-to-End Accent/Dialect Identification }

For scenarios where sufficient training data for target accents/dialects is available, a promising approach to achieve robust accent classification and embedding is to train an E2E neural network-based model. However, accent identification can prove to be more challenging compared to speaker or language identification, particularly since the degree of accent falls within a spectrum, thus transforming it into a binary classification task with fuzzy boundaries. Moreover, the task of discriminating between similar dialects poses an additional challenge, exemplified in cases such as differentiating American English (en-US) from Australian English (en-AU).

\begin{figure}
\includegraphics{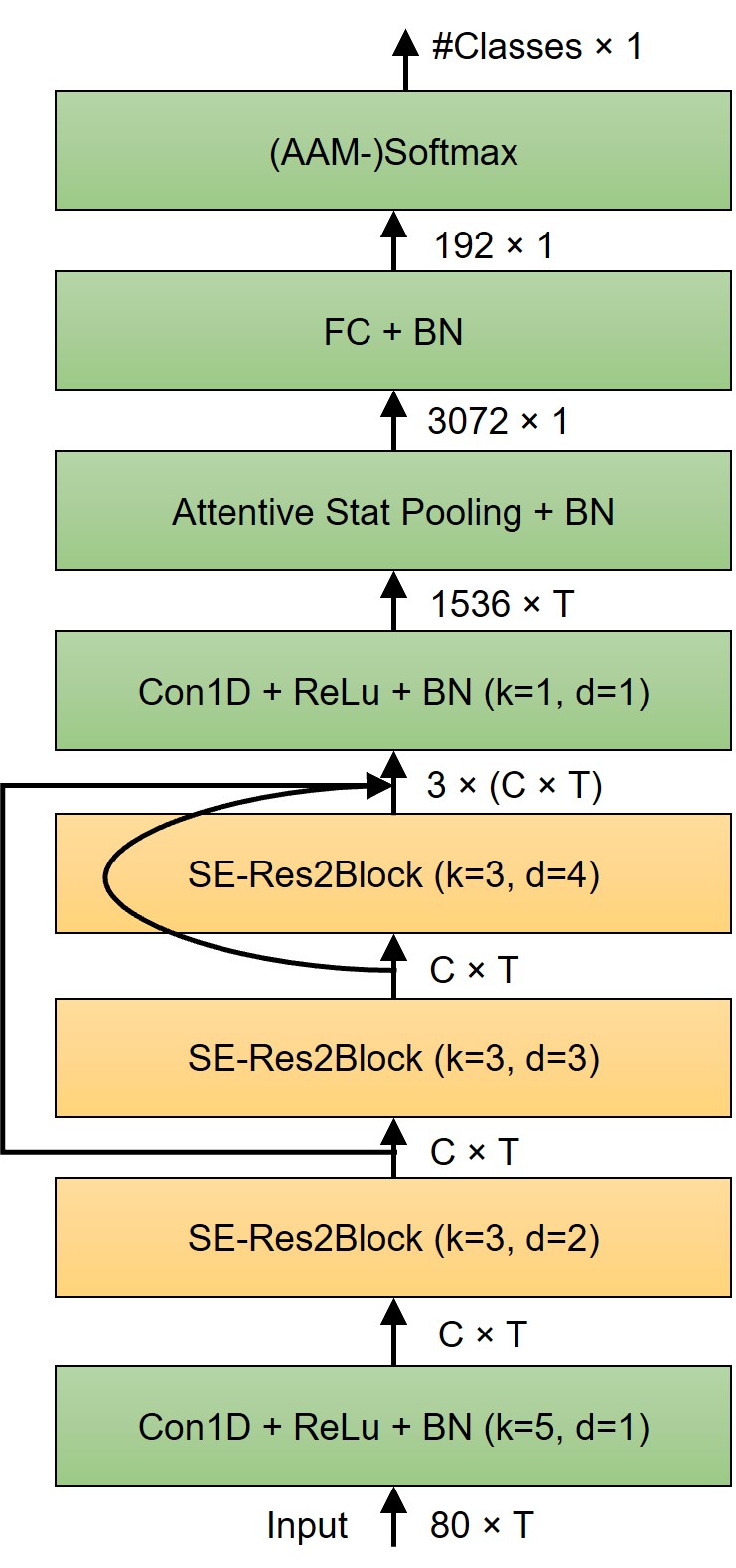}
\caption{\label{fig:ecapa}{Block diagram of the ECAPA-TDNN model~\citep{ecapa}}}
\end{figure}

Here, we use the ECAPA-TDNN~\citep{ecapa} architecture for the task of E2E accent identification. ECAPA-TDNN, shown in Fig.~\ref{fig:ecapa}, represents an improvement upon the widely used x-Vector architecture~\citep{x-vector}, utilizing 1-dim Res2-blocks with hierarchical grouped convolutions to enhance performance, while reducing the number of trainable parameters. Squeeze Excitation (SE) blocks are employed to rescale the intermediate feature maps' channels and incorporate global context information into the locally operating convolutional blocks. The outputs of the SE-Res2Blocks are then aggregated and passed through dense layers to generate effective per-frame features for the pooling layer. The pooling layer employs a channel- and context-dependent statistical approach to selectively attend to different frames per channel, allowing the model to assign higher weights to more relevant frames. The output layer also utilizes the softmax activation function to generate a distribution over the accent/dialect classes.

\subsection{\label{subsec:2:1} Pre-Trained Language Identification Model }

A language identification system is designed to automatically recognize the language of an utterance. These systems play an important role in various applications, such as multi-lingual spoken translation, multi-lingual speech recognition, diarization, and human-machine communication systems. Advanced LID systems are typically trained in an end-to-end manner using large amounts of training data containing over 100 languages. These systems process the entire utterance and extract language-discriminating information within a fixed-length embedding vector. These vectors embody extensive information regarding the input utterance. For example, the relative distance between embedding vectors representing the structure of language families, as demonstrated in earlier research~\citep{LIDaccent}. Recent studies have shown that performance of LID systems decrease significantly when evaluated with speech from non-native or regional accents \citep{LIDaccent}. This highlights that embedding-based LID models have a different representation for native and non-native speakers of a language. Here, we investigate leveraging embedding information from an advanced LID system as auxiliary features to train accent identification systems.

A pre-trained ECAPA-TDNN model is employed, which is trained on the VoxLingua107 dataset~\citep{valk2021slt} using the SpeechBrain framework~\citep{speechbrain}. Compared to the conventional ECAPA-TDNN design, our LID model incorporates additional feed-forward layers following the embedding layer and is optimized to minimize the cross-entropy loss. The VoxLingua107 is a speech dataset used to train LID models, consisting of short speech segments extracted from YouTube videos. The classification error rate of the pre-trained LID model on the development dataset of VoxLingua107 is 6.7\%.

\subsection{\label{subsec:2:2} Pre-trained Speaker Identification Model}

Speaker identification systems are designed to automatically recognize speaker's identity of speech. These systems are commonly employed for authentication in secure systems and forensic evaluations. Like LID systems, advanced SID systems are trained end-to-end using datasets encompassing over 1000 speakers. These systems typically employ pooling layers to convert variable-length utterances into compact, fixed-length speaker-characterizing embeddings~\citep{ecapa}.

For the pre-trained SID model, we leverage an ECAPA-TDNN model trained with the SpeechBrain~\citep{speechbrain} framework on the VoxCeleb~\citep{voxceleb2} training data. The VoxCeleb dataset consists of short speech segments also extracted from YouTube interviews and includes two versions, VoxCeleb1~\citep{voxceleb1} and VoxCeleb2~\citep{voxceleb2}. 
 The first version contains over 150,000 utterances from 1251 celebrities, and the second includes over a million utterances from more than 6,000 speakers. The VoxCeleb dataset is multilingual, collected from speakers of 145 different nationalities, encompassing a wide range of accents. The performance of the SID model used in this study is evaluated on the VoxCeleb1 test set (cleaned) with an equal-error-rate (EER) of 0.80\%. We hypothesize that the resulting SID model encodes accent information from speech. Therefore, we suggest leveraging the embeddings from this model as an auxiliary feature for accent/dialect identification.

\subsection{\label{subsec:2:4} Multi-Embedding Accent/Dialect Identification }

\begin{figure}
\includegraphics{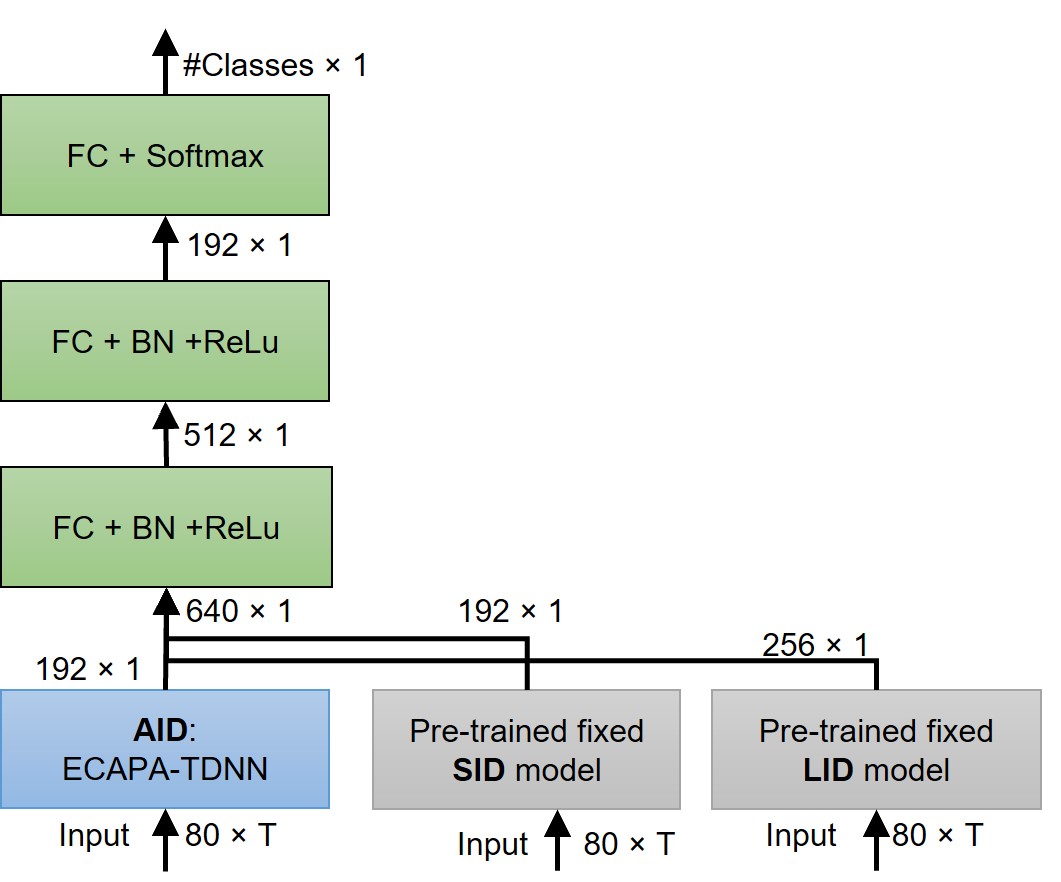}
\caption{\label{fig:me-aid}{Block diagram of the multi-embedding accent identification model. The pre-trained SID and LID models are fixed. }}
\end{figure}

To advance AID performance, we propose a multi-embedding accent identification model which leverages the information encoded by separate LID, SID, and AID models, as shown in Fig.~\ref{fig:me-aid}. 
These E2E-trained models are optimized for different tasks and trained on different datasets, leading to unique embeddings that contain distinct information for accent/dialect classification. We hypothesize that these models encode complementary information from input speech. Therefore, by concatenating the AID embedding with the LID and SID embeddings generated from pre-trained models, we aim to take advantage of the complementary information encoded by each model. The resulting combined 3-way embedding is then processed through two non-linear feed-forward layers to produce the final mixed embedding that is processed in the final classification layer. The output layer is a softmax activation function that generates a distribution over the accent/dialect classes.

\subsection{\label{subsec:CL} Leveraging Curriculum Learning (CL) for Accent/Dialect Identification}

Previous studies on LID and SID have demonstrated that classification performance significantly degrades for short speech utterances \citep{CL-SID, valk2021slt}. Accent/dialect identification on short utterances can also be particularly challenging,  since short utterances do not provide sufficient information for accurate identification, even for human listeners. Therefore, training an E2E accent classification model from scratch can be challenging when a significant number of  training utterances are very short (i.e., containing less than three words). To alleviate this issue, we adopt a curriculum learning approach \citep{CL} in order to stabilize the training process.

The critical step in curriculum learning is to divide the training data into subsets based on a corresponding difficulty criterion. Next, during training, the CL approach introduces more data sequentially in order of increasing or decreasing difficulty \citep{CL}.  This approach has been investigated for various applications including ASR~\citep{CL-ASR} and speaker recognition~\citep{CL-SID}. In~\citep{CL-ASR}, utterance length is used as the criterion. The ASR model is trained by iterating through short utterances first, considering them as easier training samples. Alternatively, in \citep{CL-SID}, the signal-to-noise ratio (SNR) of training utterances is used as the CL criterion for training a SID model. In our study, we use utterance length as the CL criterion. First, the training data is divided into $N$ subsets based on utterance length, with the last most challenging subset containing the shortest utterances. In training epoch $i$, training subsets $1, 2, ..., i$ are used for training. This approach requires only one hyperparameter, $N$, which is tuned based on the model's accuracy on the held-out validation sets. 

\section{\label{sec:3} Automatic Accentedness Assessment}
This section presents the automatic accentedness assessment systems: AID-based and ASR-based, which are detailed in Sec.~\ref{acc:1} and \ref{acc:2}, respectively.

\subsection{\label{acc:1} Accentedness Assessment Based on Accent Identification  }

AID models are optimized to accurately classify samples from different accents by maximizing the classification margin. Such training leads to an embedding space where utterances are dispersed based on their relative accent similarity. Therefore, the relative position of accented utterances with respect to en-US samples is correlated with the accentedness level of the utterances, despite the fact that other important factors, such as utterance length/duration and conversational context also impact relative accent position of utterances. Based on this, we postulate that utterances closer to en-US samples are more likely to exhibit a lower degree of accentedness, and suggest leveraging the AID model's en-US score as an objective measure. To achieve a more accurate assessment of accentedness, we suggest training a binary classification AID model specifically designed to classify en-US speech against potential target accented speech.



\subsection{\label{acc:2} Accentedness Assessment Based on Automatic Speech Recognition}


In this ASR-based accentedness estimation system, we use the recognition error-rate as an accentedness metric. The objective is to establish a strong correlation between recognition error rate and level of accentedness. Inspired by studies involving human listener recognition of accented speech we focus on two aspects of ASR systems: semantic context modeling and performance in recognizing accented speech. Previous research has demonstrated that two factors influencing human listeners' intelligibility rates for accented speech are their prior exposure to specific domain of non-native speech, and the semantic context in which the particular speech is presented. For example, a study by Bettina \citep{beinhoff2014perceiving} demonstrated that perceived intelligibility is influenced by factors such as familiarity with the relevant accent. Similarly, in \citep{kennedy2008}, it was demonstrated that listeners with more experience in accented speech have a higher intelligibility rate on L2 speech versus listeners with less experience. Furthermore, that study also revealed that more semantic context is associated with higher intelligibility rates for L2 speech for both experienced and inexperienced listeners.


%

  Current advanced ASR systems perform well for native and accented speech by pooling correctly-pronounced utterances from several accents to train a single multi-accent E2E model~\citep{domainexpansion}. However, for accentedness assessment purposes, the ASR model should actually mimic an inexperienced listener. To achieve this, we propose an ASR model trained exclusively on en-US transcribed speech. In terms of model architecture, E2E ASR techniques can generally be classified into three categories: Attention-based Encoder-Decoder (AED), Connectionist Temporal Classification (CTC), and Recurrent Neural Network Transducer (RNN-T). Among them, CTC~\citep{CTC} assumes conditional independence, meaning it does not model the inter-label dependencies. Therefore, a CTC-trained model does not rely on the semantic context to estimate the token's probability, but instead depends solely on the input signal. With this characteristic, we choose the CTC loss to train the ASR model.

 Our ASR model comprises six conformer \citep{conformer} layers. Each layer has four attention heads, and the hidden layer dimension of the feed-forward networks (FFN) is 256. The kernel size for the depthwise convolution layer in each conformer layer is set to 31. The last layer of the model has  $| \textbf{S}|$ outputs, where  $\textbf{S}=\{English\;characters,\; blank, space, ",", ?,!,.,'\}$. The $blank$ unit is a unique symbol used by CTC for calculating the loss. The final layer uses a softmax activation function to translate the model's output values into a probability distribution across the outputs. Given an input sequence $x$ and its corresponding target character sequence $t$, the CTC loss function is calculated as the negative log probability of the target sequence given the input sequence, 
\begin{equation} \label{eq:CTC1}
     J_{CTC}( x,t) = -ln P (t|x).
\end{equation}

The probability of a specific target label is calculated by summing the probabilities of all possible CTC paths that can be derived from $t$,
\begin{equation} \label{eq:CTC2}
P (t|x) = \sum_{a\in \mathcal{A}} P (a|x),
\end{equation}
where $\mathcal{A}$ represents the set of all possible CTC paths that can be expanded from $t$ by repeating output units and blank tokens. CTC assumes that the output units are conditionally independent given the input sequence. Therefore, we can compute $P (a|x) = \prod_{i=1}^{N} \mathcal{M}_{a(i)}(x) $ where $\mathcal{M}_{a(i)}(x)$ is the probability of the $a(i)$-th class produced by model $\mathcal{M}$ for input sequence $x$.

After training the acoustic model with the CTC loss function, we use a greedy algorithm (best path decoding) \cite{CTC} during the inference step to convert the sequence of probabilities generated by the model into an output  transcription. The greedy algorithm follows a two-step process: first, it selects the most likely output sequence units at each time step; next, it then uses a squash function to eliminate duplicate units. We evaluate the quality of the acoustic model's output by calculating the specific character-error-rate (CER). The model's hyperparameters (e.g., learning rate or dropout rate) are then adjusted to improve the overall CER performance on the held-out development sets.

\section{\label{sec:4} Experiments and Results}
Next, we conduct a series of experiments to evaluate performance of the accent/dialect identification and assessment methods developed in this study. First, we compare accent/dialect classification approaches with results summarized in Table \ref{table:1}. Next, we examine performance of the proposed accentedness assessment approaches.

\subsection{\label{subsec:4:1} Datasets}

The UT-CRSS-4EnglishAccent~\cite{cp1} corpus is used to train and evaluate the ASR and AID models. This corpus was collected at CRSS-UTDallas, and consists of transcribed speech from 420 speakers across four major English accents/dialects: US (en-US), Hispanic, Indian (en-IN), and Australian (en-AU). Each accent group is balanced for gender and age and has approximately 28 hours of training data, 5 hours of development data, and 5 hours of evaluation data consisting of both read and spontaneous speech.

To train the en-US ASR model, we supplement the US portion of the UT-CRSS-4EnglishAccent corpus with the 100-hour segment of the LIBRISPEECH corpus (Libri) \citep{librispeech}. This Libri subset has higher quality recordings and features speakers with accents closer to en-US English versus the rest of the corpus. For accent/dialect classification training and evaluation, we focus on the en-US, en-AU, and en-IN portions of the UT-CRSS-4EnglishAccent corpus.





\begin{figure*}

\figline{\fig{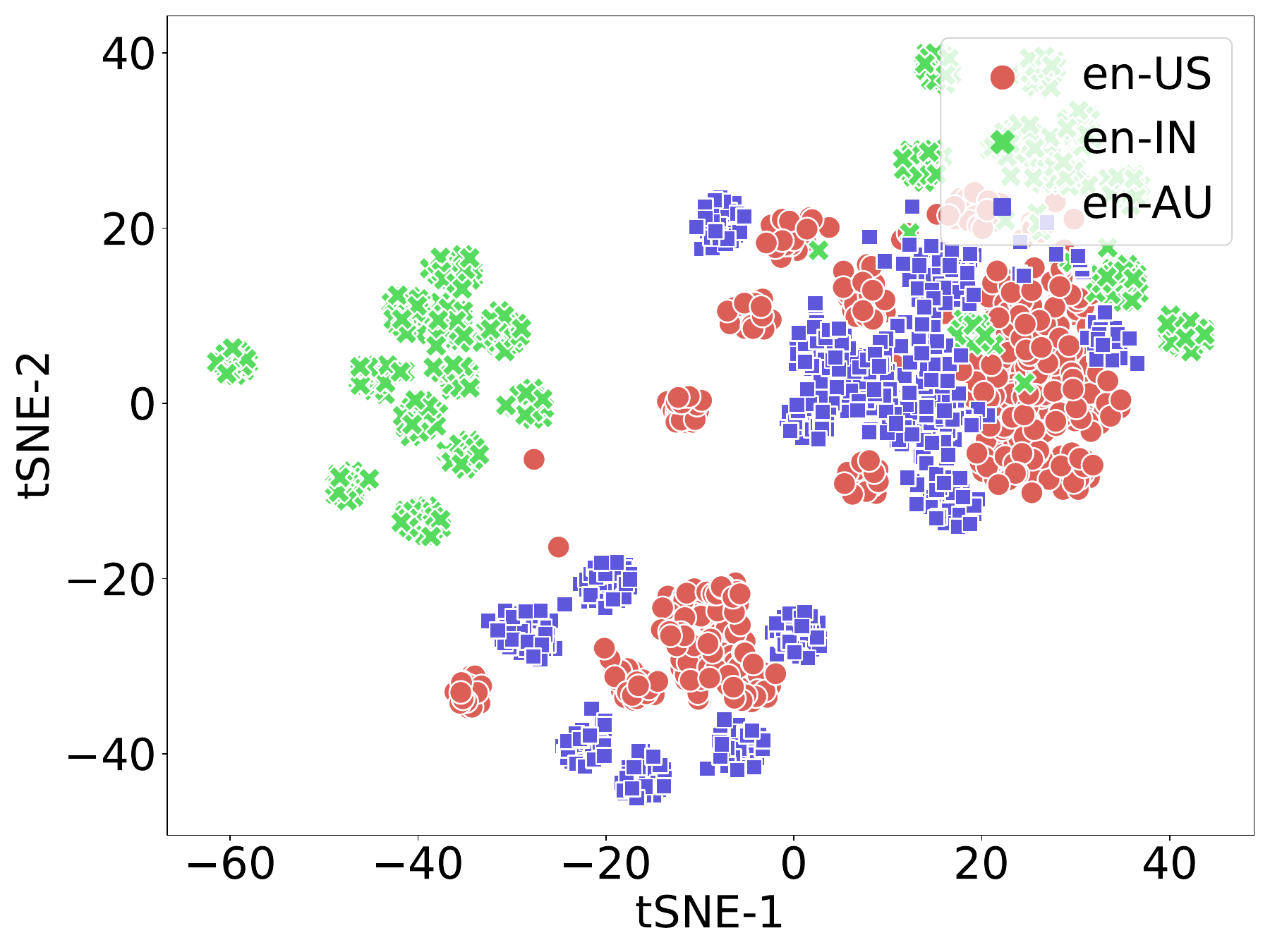}{8cm}{(a)} \label{tsne_LID}
\fig{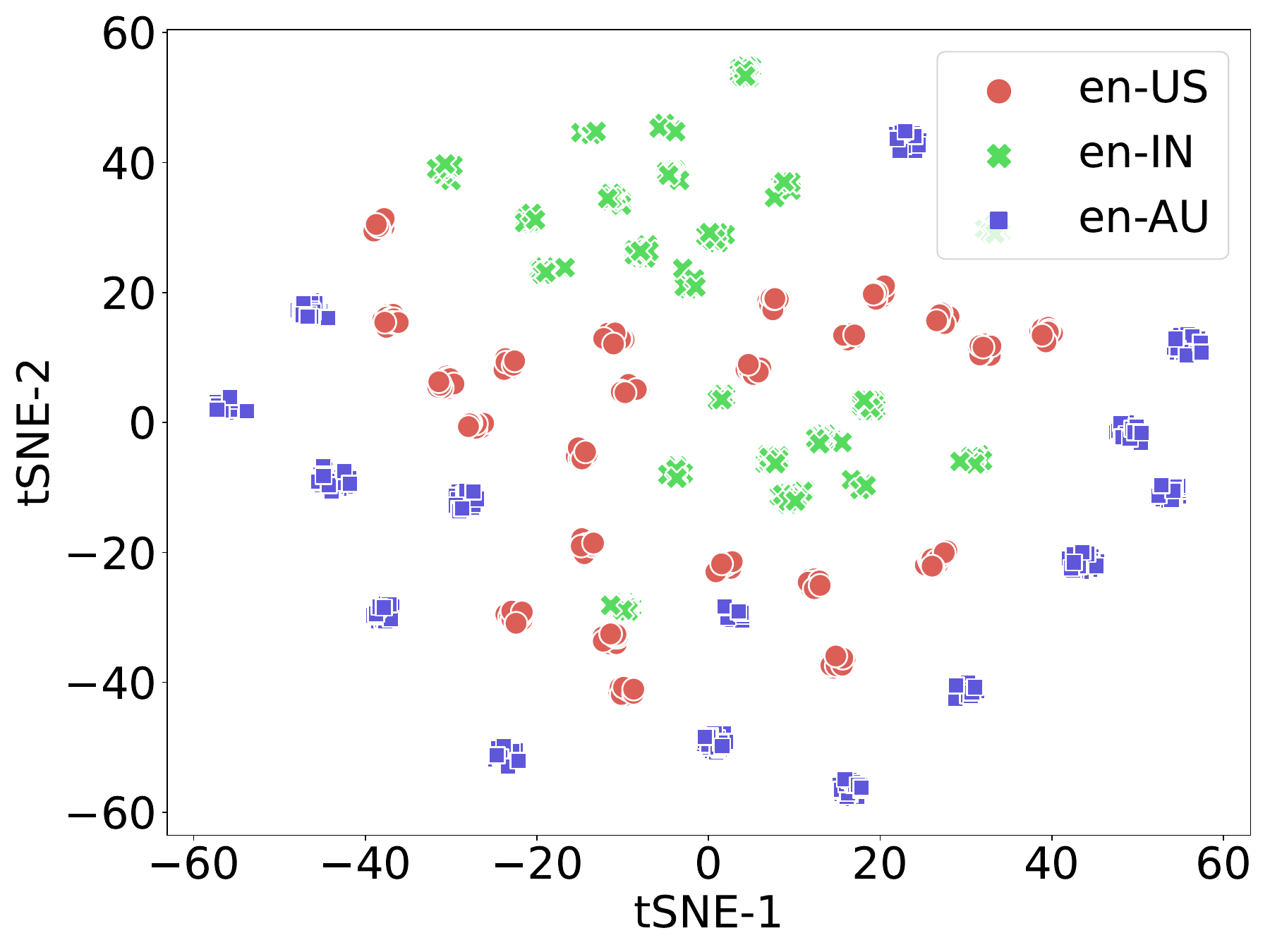}{8cm}{(b)}\label{tsne_SID}}
\figline{\fig{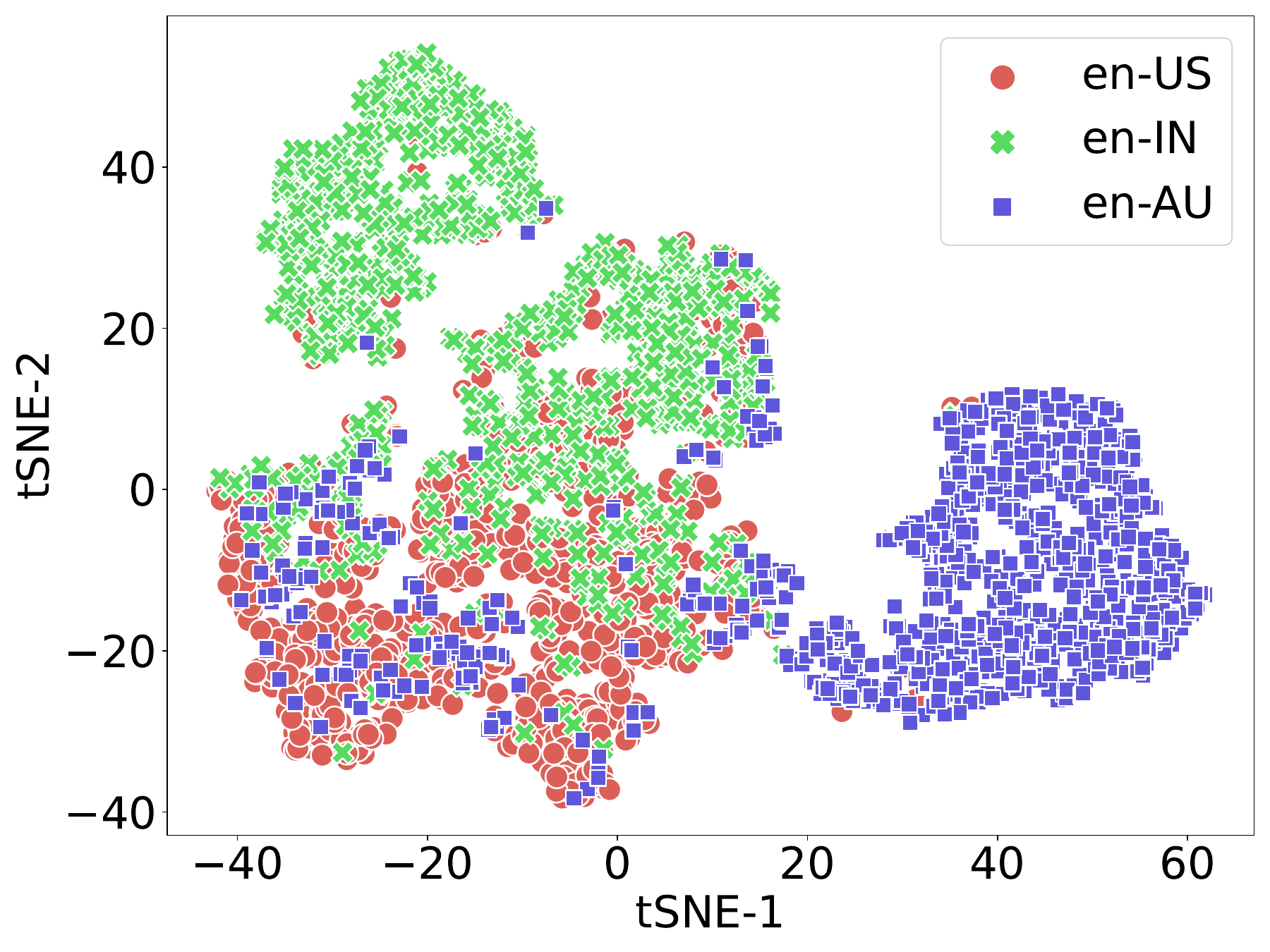}{8cm}{(c)}\label{tsne_AID}
\fig{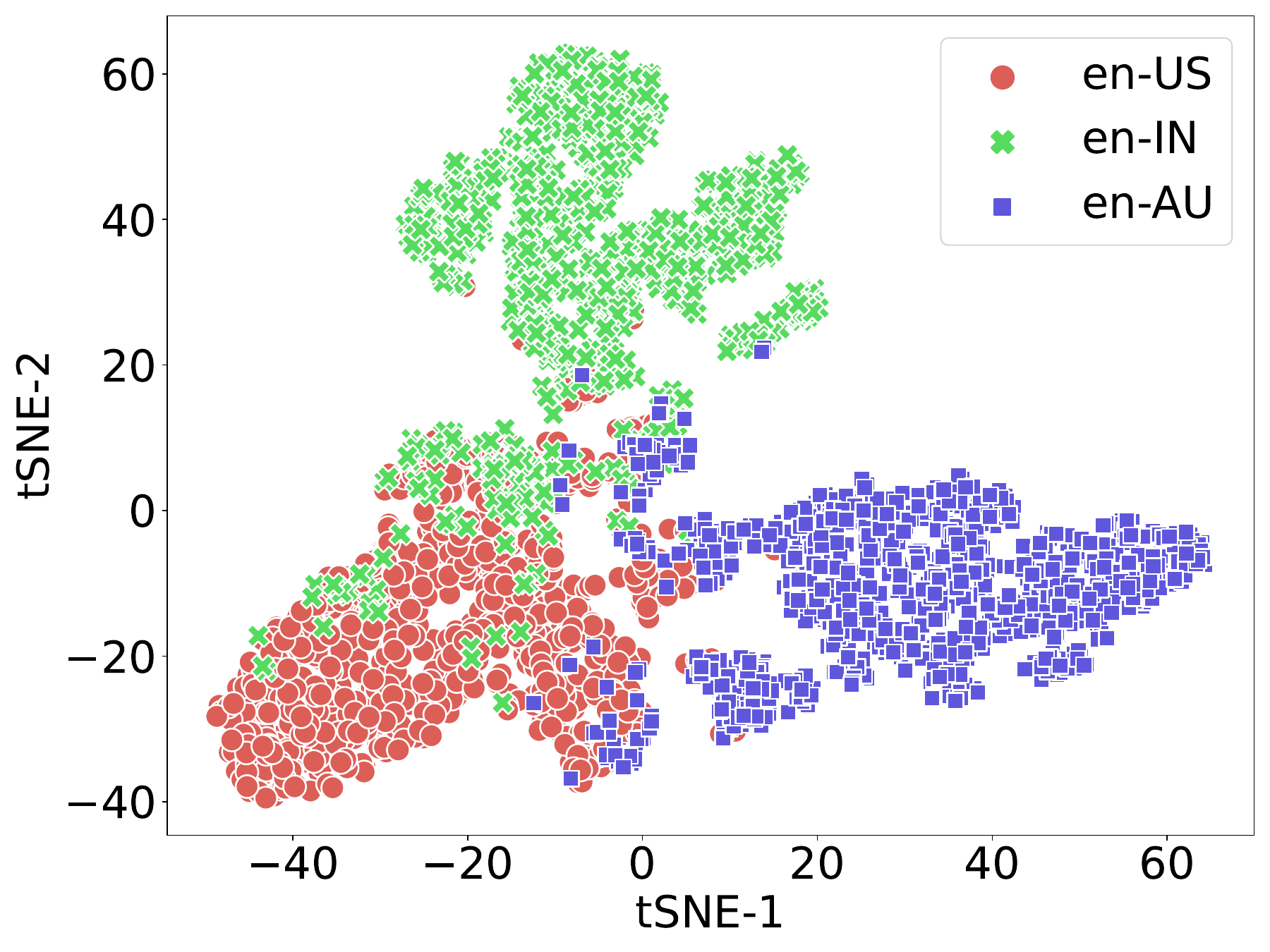}{8cm}{(d) }\label{tsne_ALL}}

\caption{\label{fig:embedding} 
Two-dimensional visualization of embeddings for en-US, en-IN, and en-AU utterances generated by pre-trained models; (a) Language identification (LID) embedding, (b) Speaker identification (SID) embedding, (c) End-to-End Accent/Dialect identification (E2E-AID) embedding, and (d) Multi-embedding AID. Each point represents an utterance, and colors correspond to different accent/dialect labels.}

\end{figure*}





\subsection{ Language and Speaker Embeddings for Accent/Dialect Identification}

\paragraph{LID embedding}~The pre-trained LID model covers 107 languages, including several commonly spoken Indian languages such as Hindi, Gujarati, Kannada, Marathi, Panjabi, Tamil, and Telugu. To assess the accuracy of the LID model for  language identification on accented speech, we examine its performance on en-IN development utterances. Out of 2401 utterances, 702 are classified as en-US, and 710 are misclassified as one of the Indian languages. This finding aligns with earlier research~\citep{LIDaccent} that has demonstrated reduced accuracy of LID models when applied to accented English speech. Furthermore, Fig.~\ref{tsne_LID} illustrates a 2-D representation of LID embeddings for three dialects/accents of English using t-SNE tool~\citep{scikit-learn}. In this LID embedding space, en-IN utterances are distinguishable using a non-linear boundary, while en-AU and en-US utterances are not as easily distinguishable.

To further investigate the effectiveness of LID embedding in classifying accents/dialects, a three-layer feed-forward neural network is trained using the fixed embeddings extracted from the en-US, en-AU, and en-IN training sets. The resulting model achieves an average accuracy of 81.3\% on the test sets, with accuracy for en-IN utterances being significantly higher at 90.8\%, as expected. These results suggest that the LID embedding effectively classifies accents/dialects, with a particular advantage for non-native accents such as en-IN.


\paragraph{SID embedding} Next, Fig.~\ref{tsne_SID} illustrates 2-D t-SNE representation of SID embedding for three dialects/accents of English. The SID embeddings for each speaker are in one cluster, reflecting the training target of the SID model. 
The boundary between dialects/accents is less clear than for speakers, however, they can be classified using non-linear boundaries with a soft margin, where accuracy is reduced. We again trained a three-layer feed-forward neural network on the SID features, following the same process as with LID features. The resulting accuracy of 75.8\% suggests that SID embeddings are less effective versus LID features for accent/dialect classification.

Comparing the 2-D representations of LID and SID embeddings (Fig.~\ref{fig:embedding}) indicates that LID features are more effective in classifying the en-IN utterances, while SID features are better at classifying the en-AU utterances from other dialects/accents. To investigate whether LID and SID models actually encode complementary information from the input signal, we trained an accent identification model that leverages both embeddings. The embeddings are concatenated and passed through a multi-layer FFN. The resulting model achieved an accuracy of 84.1\%, supporting the notion that LID and SID features contain complementary information  that can be leveraged for the task of accent/dialect identification.

\begin{table}
\centering
\caption{Comparison of model performance on accent/dialect identification task for en-US, en-IN, and en-AU. The table reports the model architecture, input features, and whether curriculum learning (CL) is utilized. The average accuracy achieved on the test sets is presented for each model.}
\vskip3pt
\begin{tabular}{cccc}
\hline\hline
 Model&Input Features&CL&Accuracy\\
\hline
M1:\quad \quad FFN&LID & \xmark & 81.3\% \\
M2:\quad \quad FFN&SID & \xmark & 75.8\% \\
M3:\quad \quad FFN&LID \& SID & \xmark & 84.1\% \\
M4:E2E-AID&MFB & \xmark & 70.2\% \\
M5:E2E-AID &MFB & \checkmark & 80.5\% \\

\bf{M6:ME-AID} & MFB \& SID \& LID & \checkmark & \bf{87.4\%} \\

\hline\hline
\label{table:1}
\end{tabular}
\end{table}

\subsection{ E2E Accent Identification Results}

The E2E accent identification (E2E-AID) model uses 80-dim log Mel filterbank (MFB) energies extracted from 25ms windows with a 10ms frame shift as input features. Features are mean normalized per input. Models are trained using a cyclical learning rate strategy, where the rate oscillates between 1e-8 and 1e-2 using the triangular2 schedule \citep{cyclical}.  An Adam optimizer~\citep{kingma2014adam} is employed to train or adapt the models. Training is done for 10 epochs with batches of 32 utterances. To prevent overfitting, a weight decay of 2e-5 is applied to all weights in the model. The final fully-connected layer has three nodes, with a softmax activation used to map the model's logits to a probability distribution over an output set. Finally, the bottleneck size in both the SE-Block and the attention module is fixed at 128.

The E2E-AID model is trained on utterances from the en-US, en-AU, and en-IN dialects of the UT-CRSS-4EnglishAccent corpus. The accent accuracy achieved with this model on the test sets is 70.2\%, which is lower than the performance of AID models trained with LID and SID features. Here, we apply the curriculum learning approach to enhance the E2E training stability and performance, as described in Sec.~\ref{subsec:CL}. This curriculum learning approach involves dividing the training data into four subsets based on utterance length, where each subset contains increasingly shorter utterances (e.g., for AID, it is assumed that longer duration utterances are easier versus short durations due to phoneme and context structures). The first subset contains utterances with more than 20 words; the second contains utterances with 11-20 words; the third contains utterances with 3-10 words; the last subset contains utterances with 1-2 words. The model gradually learns to better capture the phonetic and prosodic variations across different accents by training on progressively shorter utterances. The final model was selected based on its accuracy on held-out validation sets. The resulting model achieved an accuracy of 80.5\% on the test sets, which is +10.3\% better than the E2E-AID model without CL. This improvement illustrates the effectiveness of the proposed CL method for E2E AID modeling. A 2-dim representation of the improved E2E-AID embedding is illustrated in Fig.~\ref{tsne_AID}. Based on this plot, we can see larger and more distinct clusters that reflect accents/dialects compared to the corresponding LID and SID plots (Fig.~\ref{tsne_LID} and Fig.~\ref{tsne_SID}). However, it is also apparent that some utterances of different accents still overlap, indicating the potential for further improvement of this AID model.

To train the multi-embedding AID model, shown earlier in Fig.~\ref{fig:me-aid}, we utilized the pre-trained LID and SID models to extract fixed-length embeddings for each utterance. We also employed an E2E-AID model corresponding to model M5 in Table~\ref{table:1} to extract AID embeddings. Two methods were explored to integrate the E2E-AID model within the ME-AID architecture. First, in a fine-tuning approach, the pre-trained E2E-AID parameters and the feed-forward layers are adapted. Alternatively,  the pre-trained E2E-AID model is not further adapted, while only the feed-forward layers are trained in this stage. Experiments show that the latter approach results in superior performance. Furthermore, the same curriculum learning subsets used in  E2E-AID training are also used for the ME-AID model. The resulting model (referred to as model M6 in Table~\ref{table:1}) achieves an accuracy of 87.4\%, outperforming all other approaches in this study. This superior performance supports the hypothesis that combining different embeddings in our ME-AID model leads to a more robust and accurate overall accent classification system. A t-SNE 2-dim representation of the utterances in the multi-embedding space (see Fig.~\ref{tsne_ALL}) clearly demonstrates more distinct boundaries between accents/dialects versus earlier evaluations (Fig.~\ref{tsne_LID}-\ref{tsne_AID}).

\begin{table}
\centering
\caption{Accuracy of the ME-AID model (M6) on test subsets based on the length of utterance. From Table~\ref{table:1}, overall the ME-AID performance is 87.4\%.}
\vskip3pt
\label{table:2}
\begin{tabular}{cccccc}
\hline\hline
 Model&1-2 words&3-10 words&11-20 words& 21-200 words\\
 &1929 utts&2238 utts&1456 utts&896 utts\\
\hline
M6&80.6\%  & 87.8\%  & 91.7\% & 94.2\%  \\

\hline\hline
\end{tabular}
\end{table}

  Next, Table~\ref{table:2}  presents performance of the ME-AID model (corresponding to model M6 in Table~\ref{table:1}), on different subsets of the test sets, based on utterance test length/duration. Test sets are divided into four groups: 1-2 words, 3-10 words, 11-20 words, and 21-200 words. The number of utterances in each subset is also presented in the table. Results demonstrate that the model's accuracy improves with utterance duration length, with an accuracy of 80.6\%, 87.8\%, 91.7\%, and 94.2\% for the first, second, third, and fourth subsets, respectively, compared to overall performance rate of 87.4\%. These results highlight the significant impact of utterance length for accurate accent/dialect identification.

\begin{table}[ht]
\caption{ Confusion matrix of the ME-AID model. The rows and columns represent the true accent labels and the predicted accent labels, respectively. For each row, the values in the cells represent the percentage of utterances that were classified as the corresponding predicted accent label.}
\label{table:3}
\vskip3pt
\begin{tabular}{ccccc}
\hline\hline
 &en-US&en-IN&en-AU\\
\hline
en-US & 85.1\%  & 10.4\% & 4.5\% \\
en-IN & 8.8\%  & 90.0\% & 1.2\% \\
en-AU & 10.5\%  & 3.3\% & 86.2\% \\

\hline\hline
\end{tabular}
\end{table}

\begin{figure*}

\figline{\fig{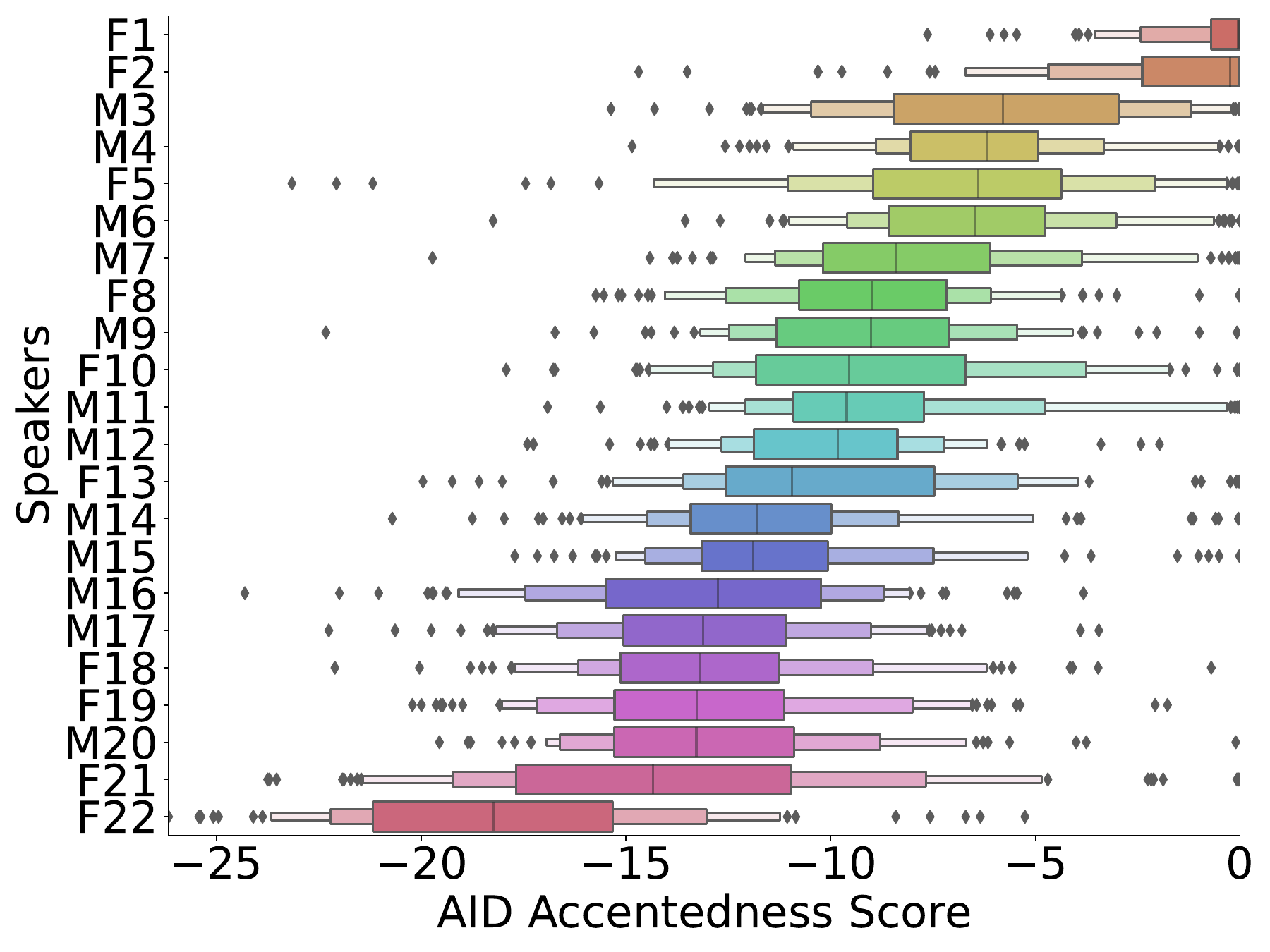}{8cm}{(a) AID-based accentedness scores }
\fig{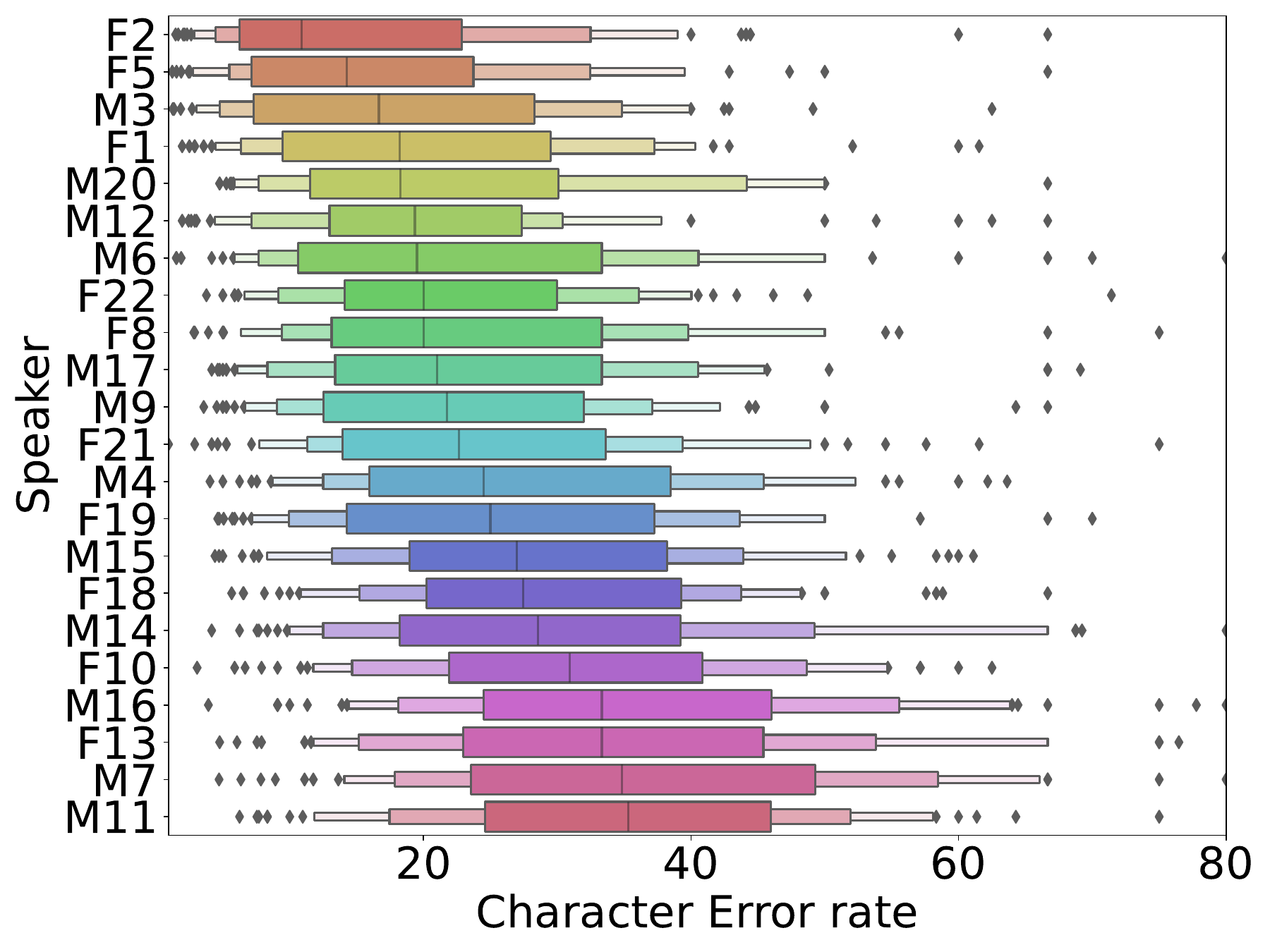}{8cm}{(b) ASR-based accentedness scores }}

\caption{ \label{fig:boxplot}   Box plot comparing the distribution of accentedness scores for the en-IN test set using two different methods: (a) AID-based approach, where the log\_softmax of the en-US output is used to estimate the level of accentedness, with higher scores indicating a lower estimated level of accentedness; (b) en-US ASR model-based approach, where the character-error-rate of the ASR model is used as an accentedness metric, with higher scores indicating a higher estimated level of accentedness. The x-axis shows estimated scores, and the y-axis represents ranked speakers based on the median of their accented scores. Speakers are numbered based on their AID rank, with prefix "F" for females and "M" for males. }

\end{figure*}

The results in Table~\ref{table:3} illustrate the confusion matrix performance of the ME-AID system on different accents. The model achieves high accuracy in classifying en-IN, with 90.0\% accuracy. However, accuracy for en-AU and en-US accents is relatively lower, at 86.2\% and 85.1\%, respectively. These results suggest that the model performs well in differentiating en-IN from other accents, but requires further improvement in classifying English dialects en-AU and en-US. Moreover, the model tends to misclassify a relatively high percentage of en-IN and en-AU accents as en-US.  This may be attributed to variations within the en-US dataset and warrants further investigation.


\subsection{ Accentedness Assessment Results}

In this section, we evaluate performance of accentedness assessment systems based on the accent/dialect identification system and automatic speech recognition systems presented in Sec.~\ref{acc:1} and \ref{acc:2}, respectively. We use the en-IN test set from the UT-CRSS-4EnglishAccent corpus as our non-native accent evaluation dataset to conduct this evaluation.

The ME-AID model is leveraged as the AID system since it demonstrated superior classification performance compared to other approaches. However, we retrained the model specifically for classifying en-US and en-IN utterances to reduce multi-class confusion. The resulting model achieved an overall test set classification accuracy of 91.2\%. To estimate the accentedness level of each utterance, we employ the log\_softmax of the en-US output of the model, where higher numerical scores correspond to lower estimated levels of speech accentedness.

 We compute an accentedness score of all utterances from a given speaker and order the speakers based on the median of their scores. Overall, this allows for a speaker assessment versus an individual utterance assessment. To evaluate the effectiveness of the estimated scores,  the accentedness rankings of speakers determined by the AID and ASR models are compared. In the following section, we further examine the correlation of these objective scores versus subjective listener scores based on a separate human perception assessment.

Fig.~\ref{fig:boxplot} presents an enhanced box plot that visualizes the distribution of AID accentedness scores for the en-IN test set. The x-axis represents the AID accentedness scores, while the y-axis represents the ranked speakers based on their median accentedness scores. Utilizing the seaborn library~\citep{seaborn}, the enhanced box plot provides a more detailed representation of the data, including the median, quartiles, and outliers. The en-IN test set comprises 22 M/F speakers, with speaker F1 ranked as the least accented, and speaker F22 as the most accented, according to the AID accentedness estimation approach. This visualization offers a clear ranked understanding of the distribution of accentedness scores among speakers in the test set. Notably, some speakers in the plot have a higher variance in their accentedness scores, indicated by the larger spread of their box plots. This could be attributed to the length of their evaluation utterances, as we have previously established that the ME-AID model is more robust and accurate for longer-duration utterances. Additionally, these speakers might have exhibited a wider range of accent variations within their speech, contributing to the increased variance in their scores.

Next, we evaluate the effectiveness of using the ASR model to assess accentedness of non-native speech. Our E2E CTC-based ASR model was trained using 80-dim log Mel filterbank energies extracted from 25ms windows with a 10ms frame shift as input features. As noted earlier, all features are mean-normalized per input. To optimize training, we employ the ReduceLROnPlateau scheduler in PyTorch, which automatically adjusts the learning rate when the model's progress on the validation loss plateaus. The initial learning rate is set to 0.001,  and an Adam optimizer is employed to train the model. The training was performed for 40 epochs with batches of 16 utterances. The resulting model achieved a character error rate (CER) of 15\% on the en-US test set, which is from the same accent as the train data. However, the CER for en-IN is 26\%, which is significantly higher. We attribute this performance degradation to the phonetic mismatch between native and non-native speakers, as the acoustic recording conditions were constant for all accents in the UT-CRSS-4EnglishAccent corpus.

 Fig.~\ref{fig:boxplot}(b) presents a box plot illustrating the distribution of ASR-based accentedness scores for the 22 speakers in the en-IN test set. We computed CER of all utterances from a given speaker, then ranked the speakers in ascending order based on median CER of their utterances. The results indicate a strong correlation between the estimated accent level obtained from both the AID and ASR models. The speakers that rank highest in AID scores, such as F2, F5, M3, and F1, also rank the highest based on ASR scores. This suggests that the AID and ASR models are consistent in their estimation of accentedness and offer complementary viewpoints. 


To further evaluate the relationship between accentedness scores obtained from the ASR-based and AID-based approaches, we calculate the Pearson correlation between them on the en-IN test set, using the median value of each speaker's scores. The resulting Pearson coefficient of -0.402 and p-value of 0.06 indicates a moderate linear correlation between the two scoring methods. However, limiting the analysis to include utterances with more than 20 transcription words increases the correlation coefficient to -0.57, with a p-value of 0.005, indicating a stronger correlation. This finding suggests that the length of the utterance plays a significant role in accurately estimating accentedness level.

The results presented in this section therefore demonstrate the potential of leveraging ASR and AID models for assessing accentedness in non-native speakers. The strong correlation between scores obtained from the two models indicates their consistency and accuracy in estimating accentedness.







\subsection{Human Perception Scores}

In this last section, we investigate the correlation between the ASR/AID-based accentedness scores and subjective listening scores based on human perception, to assess these models' potential for automatic assessment of accentedness in non-native speech. We utilize subjective scores collected for 20 en-IN speakers in the UT-CRSS-4EnglishAccent corpus~\citep{yang2022improving}. The accentedness of sampled utterances was rated on a 1-9 scale by 17 human listeners, where 1 represents a heavy accent, and 9 essentially native. Each speaker received at least 30 ratings from different raters. The scores are aggregated to obtain a per-speaker accentedness score.   To ensure unbiased ratings, raters were only presented with the audio of the second language without any transcription or additional information. Raters were graduate students at Northern Arizona University with two or more years of experience in second language pronunciation teaching or research (e.g., more experienced in speech assessment versus general listeners).

\begin{figure}
\includegraphics[width=\reprintcolumnwidth]{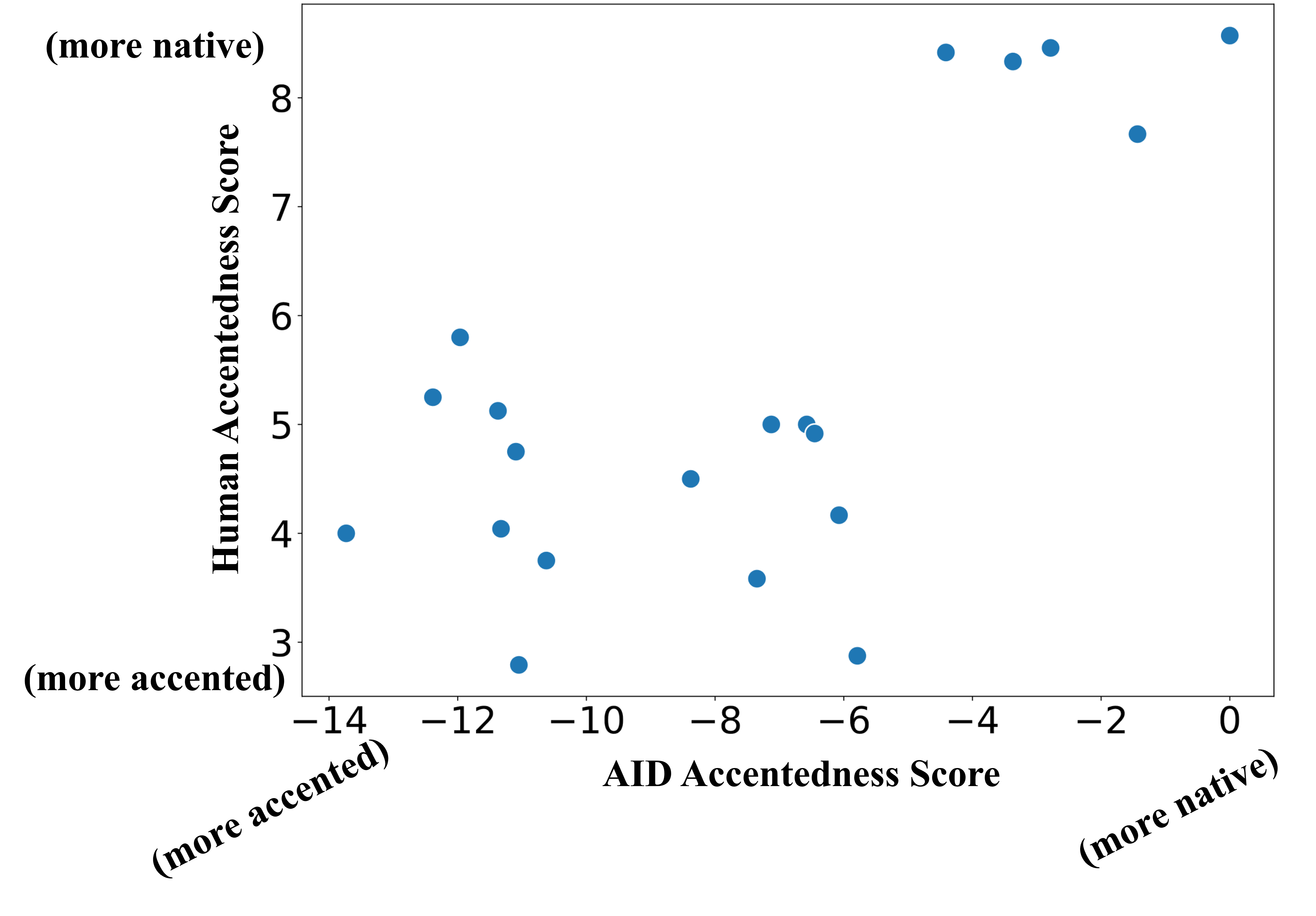} 
\caption{\label{fig:corr1}{AID-based Accentedness Score versus Human Accentedness Score. Pearson Correlation = 0.68 $(p  = 1e-3).$ }}
\end{figure}

Fig.~\ref{fig:corr1} illustrates the correlation between AID-based accentedness scores and human perception listener accentedness scores. The x-axis represents the automated AID scores, with higher scores indicating a lower estimated level of accentedness, while the y-axis represents the human perception scores, with higher scores indicating a lower level of accentedness. For specific speakers (e.g., five speakers in upper right location of the figure), the average human score is higher than the rest, and notably, the AID-based score for these same speakers is also higher. This suggests that the AID model accurately predicts the level of accentedness for these speakers as perceived by human listeners.

\begin{figure}
\includegraphics[width=\reprintcolumnwidth]{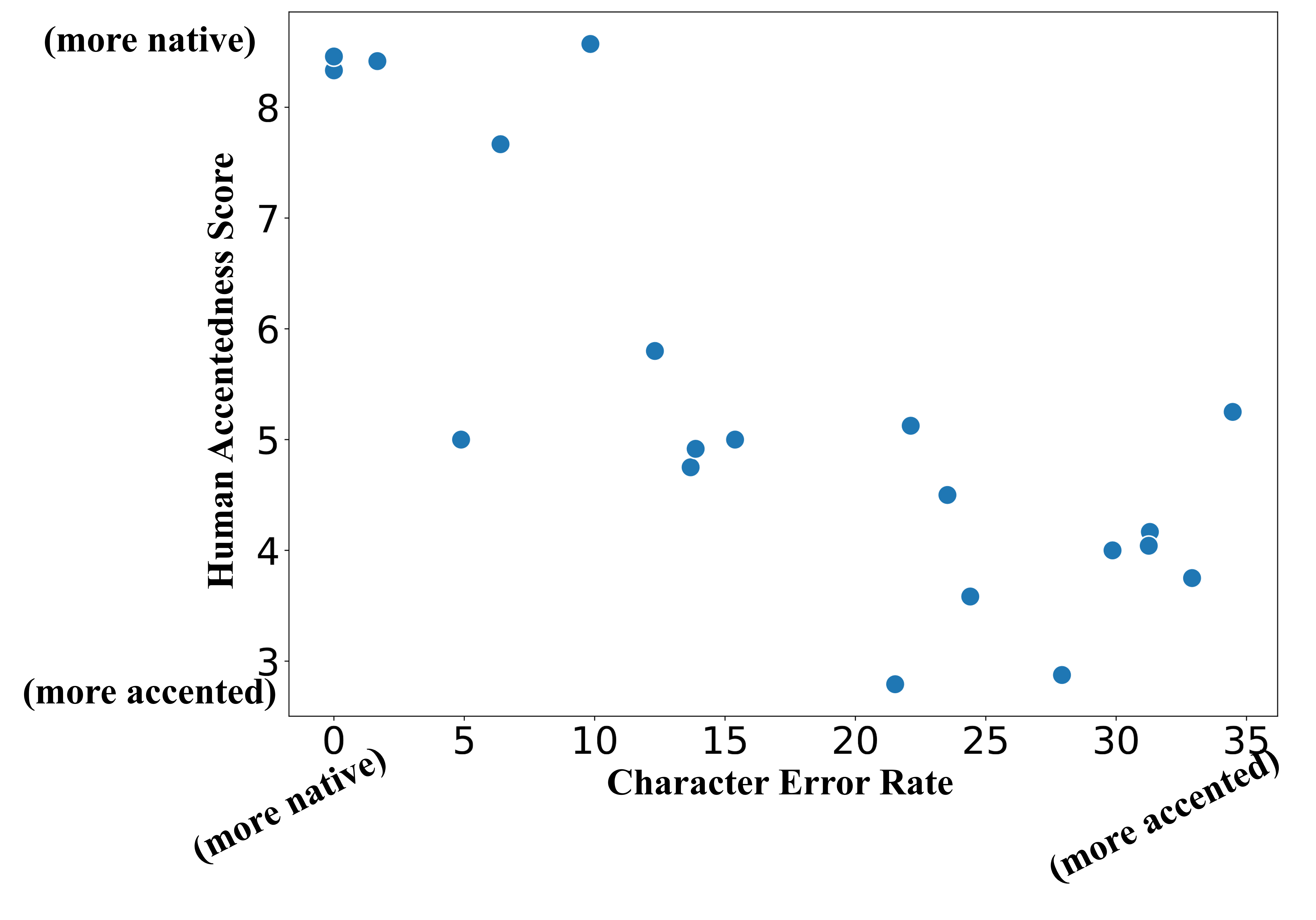} 
\caption{\label{fig:corr2}{Character Error Rate versus Human Accentedness Score. Pearson Correlation = -0.78 $(p  = 3.4e-5).$ }}
\end{figure}

Based on Fig.~\ref{fig:corr1}, the correlation between the AID scores and human perception scores is weaker for speakers with higher accent levels. This can be attributed to the inherent differences between subjective and objective scores (e.g., human perception may not align accent along a linear scale). Human factors such as familiarity with the accented L1 versus L2 languages, speech intelligibility, and semantic context can influence individual human perception, but they may not impact the automatic AID model's predictions. Furthermore, the automatic AID model is not explicitly optimized for predicting accentedness scores, and instead performs predictions based on the distance of an utterance from the en-US classification boundary in the embedding space, thus performing better for utterances that are closer to the boundary. However, despite these differences, the Pearson coefficient between AID-based and subjective scores is 0.68, with a p-value of 1-e3, indicating a moderate to strong correlation between the two scores. This further supports the reliability and validity of automatic AID-based scores to measure accentedness in non-native speech.


The correlation between ASR-based accentedness scores based on Character-Error-Rate (CER) and human perception accentedness scores is illustrated in Fig.~\ref{fig:corr2}. The ASR-based scores are computed as the average CER for each speaker. Speakers can be clustered into two groups; speakers with an average subjective score larger than 7 (lowest accentedness perceived by human listeners), and speakers with an average subjective score below 6 (more accented, non-native). For the first cluster, the proposed ASR model (trained with en-US utterances) demonstrates a small average CER, which is consistent for the performance range obtained with en-US speakers. For the second cluster, a strong linear correlation between the ASR-based scores and subjective scores is observed. This indicates that the ASR model is sensitive to speaker's accentedness level, as reflected in CER. Now assessing all speakers, the computed Pearson coefficient between the ASR-based scores and subjective scores is -0.78, with a p-value of 3.4-e5, confirming a strong linear correlation between the two scores.



In conclusion, our findings suggest that ASR-based scores have a higher correlation with human-perceived accentedness compared to the calculated AID-based scores. This is likely because the ASR model is specifically trained to recognize native en-US speech and is, therefore, more tuned to variations in accentedness from alternative phoneme/word pronunciation stand point. However, the automatic AID-based scoring has the advantage of not requiring accurate ground truth transcription knowledge to calculate the accentedness score. These findings provide additional support for using both AID-based and ASR-based scores as measures of accentedness in non-native speech. Taken together, they can provide a more comprehensive and nuanced assessment of accentedness in non-native speech.




\section{\label{sec:5}Conclusion}

This study has investigated methods for accent/dialect classification and accentedness assessment. The accent/dialect identification experiments were conducted using data from the UT-CRSS-4EnglishAccent corpus (en-US, en-AU, and en-IN). We demonstrated that utilizing advanced pre-trained language identification (LID) and speaker identification (SID) embeddings can effectively improve performance of accent/dialect identification. The LID-based model achieved   81.3\% accuracy and was particularly effective at classifying non-native speech, achieving 90.8\% accuracy for en-IN. To leverage training data, we also trained an end-to-end accent/dialect identification model (E2E-AID) which achieved 70.2\% accuracy, and improved this to 80.5\% using curriculum learning (CL). Our proposed multi-embedding accent identification (ME-AID) model, which leverages LID, SID, and E2E-AID embeddings, achieved the highest overall classification accuracy of 87.4\%. In addition, the results demonstrate that ME-AID performance improves with increasing utterance duration, highlighting the impact of utterance phoneme diversity on accent/dialect identification accuracy.

We also investigated leveraging the two solutions for accent/dialect identification system and automatic speech recognition for the core task of accentedness estimation. We suggested training a two-class ME-AID model classifying en-US and en-IN utterances for the purpose of accentedness assessment for speakers/utterances. The log\_softmax of the en-US output of the model was used to estimate the accentedness score of each utterance. For ASR-based accentedness estimation, we proposed using an end-to-end CTC ASR model trained exclusively with speech from native en-US speakers. The recognition character-error-rate (CER) of the ASR model was used as the accentedness metric. Automatic accentedness assessment methods were evaluated on the en-IN test set. The results demonstrated that the estimated accent level score obtained from AID and ASR models is highly correlated. The correlation between the AID and ASR-based accentedness estimates was moderate (Pearson coefficient of -0.402), but improved to a more robust correlation (Pearson coefficient of -0.57) when analyzing longer utterances.

Additionally, we examined the correlation between accentedness scores generated by ASR and AID systems and subjective scores based on human perception. This was explored to assess the models' potential for automatic assessment of accentedness in non-native speech. The subjective scores were collected for 20 en-IN speakers from the UT-CRSS-4EnglishAccent corpus.   The ASR-based scores demonstrated a stronger correlation with the human-perceived accentedness scores versus the AID-based scores. The computed Pearson coefficient between AID-based and subjective scores was 0.68, while the correlation between ASR-based and subjective scores was -0.78. The stronger correlation of ASR scores can be attributed to the ASR model being specifically trained to recognize speech and effectively leverage mispronunciation information. Finally, AID-based scoring does not require accurate ground truth transcription to estimate the accentedness score, and can provide a neutral quantitative assessment of utterance/speaker accent level.  
  
This study has therefore provided evidence for the reliability and validity of using both AID-based and ASR-based systems to create scores as measures of accentedness in non-native speech. Together, they can provide a more comprehensive and diagnostic assessment of accentedness in non-native speech.

\begin{acknowledgments}
This research was supported by the University of Texas at Dallas
from the Distinguished University Chair in Telecommunications Engineering
held by John H.L. Hansen.
\end{acknowledgments}


\bibliography{allbib}

\begin{thebibliography}{41}
\def\enquote#1{``#1,''}
\def\plainquote#1{``#1''}
\expandafter\ifx\csname natexlab\endcsname\relax\def\natexlab#1{#1}\fi
\providecommand{\dourl}[1]{\href{http://#1}{\nolinkurl{#1}}}
\providecommand{\bibinfo}[2]{#2}
\providecommand{\noopsort}[1]{}
\providecommand{\switchargs}[2]{#2#1}
  \def\eatspace #1{#1}

\bibitem[{Amodei \emph{et~al.}(2016)Amodei, Ananthanarayanan, Anubhai, Bai, Battenberg, Case, Casper, Catanzaro, Cheng, Chen \emph{et~al.}}]{CL-ASR}
\bibinfo{author}{Amodei, D.}, \bibinfo{author}{Ananthanarayanan, S.}, \bibinfo{author}{Anubhai, R.}, \bibinfo{author}{Bai, J.}, \bibinfo{author}{Battenberg, E.}, \bibinfo{author}{Case, C.}, \bibinfo{author}{Casper, J.}, \bibinfo{author}{Catanzaro, B.}, \bibinfo{author}{Cheng, Q.}, \bibinfo{author}{Chen, G.} \emph{et~al.} (\textbf{\bibinfo{year}{2016}}). \enquote{\bibinfo{title}{Deep speech 2: End-to-end speech recognition in english and mandarin}} \bibinfo{journal}{Inter. Conf. on Machine Learning (ICML),} \bibinfo{pages}{pp. 173--182}.

\bibitem[{Arslan and Hansen(1996)}]{arslan1996language}
\bibinfo{author}{Arslan, L.~M.},  and \bibinfo{author}{Hansen, J.~H.} (\textbf{\bibinfo{year}{1996}}). \enquote{\bibinfo{title}{Language accent classification in american english}} \bibinfo{journal}{Speech Communication} \textbf{18}(4), \bibinfo{pages}{353--367}.

\bibitem[{Arslan and Hansen(1997)}]{arslan1997frequency}
\bibinfo{author}{Arslan, L.~M.},  and \bibinfo{author}{Hansen, J. H.~L.} (\textbf{\bibinfo{year}{1997}}). \enquote{\bibinfo{title}{Frequency characteristics of foreign accented speech}} \bibinfo{journal}{IEEE Inter. Conf. Acoustics, Speech, and Signal Proc. (ICASSP)} \textbf{2}, \bibinfo{pages}{pp. 1123--1126}, \dodoi{10.1109/ICASSP.1997.596139}.

\bibitem[{Beinhoff(2014)}]{beinhoff2014perceiving}
\bibinfo{author}{Beinhoff, B.} (\textbf{\bibinfo{year}{2014}}). \enquote{\bibinfo{title}{Perceiving intelligibility and accentedness in non-native speech: A look at proficiency levels}} \bibinfo{journal}{Inter. Symposium on Acquisition of Second Language Speech} \textbf{5}, \bibinfo{pages}{pp. 58--72}.

\bibitem[{Bengio \emph{et~al.}(2009)Bengio, Louradour, Collobert, and Weston}]{CL}
\bibinfo{author}{Bengio, Y.}, \bibinfo{author}{Louradour, J.}, \bibinfo{author}{Collobert, R.},  and \bibinfo{author}{Weston, J.} (\textbf{\bibinfo{year}{2009}}). \enquote{\bibinfo{title}{Curriculum learning}} \bibinfo{journal}{Inter. Conf. on Machine Learning (ICML),} \bibinfo{pages}{pp. 41--48}, \dodoi{10.1145/1553374.1553380}.

\bibitem[{Chung \emph{et~al.}(2018)Chung, Nagrani, and Zisserman}]{voxceleb2}
\bibinfo{author}{Chung, J.~S.}, \bibinfo{author}{Nagrani, A.},  and \bibinfo{author}{Zisserman, A.} (\textbf{\bibinfo{year}{2018}}). \enquote{\bibinfo{title}{Voxceleb2: Deep speaker recognition}} \bibinfo{journal}{ISCA Interspeech,} \bibinfo{pages}{pp. 1086--1090}, \dodoi{10.21437/interspeech.2018-1929}.

\bibitem[{Deshpande \emph{et~al.}(2005)Deshpande, Chikkerur, and Govindaraju}]{AID2005}
\bibinfo{author}{Deshpande, S.}, \bibinfo{author}{Chikkerur, S.},  and \bibinfo{author}{Govindaraju, V.} (\textbf{\bibinfo{year}{2005}}). \enquote{\bibinfo{title}{Accent classification in speech}} \bibinfo{journal}{IEEE Workshop on Automatic Identification Advanced Technologies (AutoID'05),} \bibinfo{pages}{pp. 139--143}, \dodoi{10.1109/AUTOID.2005.10}.

\bibitem[{Desplanques \emph{et~al.}(2020)Desplanques, Thienpondt, and Demuynck}]{ecapa}
\bibinfo{author}{Desplanques, B.}, \bibinfo{author}{Thienpondt, J.},  and \bibinfo{author}{Demuynck, K.} (\textbf{\bibinfo{year}{2020}}). \enquote{\bibinfo{title}{{ECAPA-TDNN}: Emphasized channel attention, propagation and aggregation in tdnn based speaker verification}} \bibinfo{journal}{ISCA Interspeech,} \bibinfo{pages}{pp. 3830--3834}, \dodoi{10.21437/interspeech.2020-2650}.

\bibitem[{Ferragne and Pellegrino(2010)}]{ferragne2010formant}
\bibinfo{author}{Ferragne, E.},  and \bibinfo{author}{Pellegrino, F.} (\textbf{\bibinfo{year}{2010}}). \enquote{\bibinfo{title}{Formant frequencies of vowels in 13 accents of the {B}ritish isles}} \bibinfo{journal}{Journal of the Inter. Phonetic Assoc.} \textbf{40}(1), \bibinfo{pages}{pp. 1--34}, \dodoi{10.1017/s0025100309990247}.

\bibitem[{Franco \emph{et~al.}(1999)Franco, Neumeyer, Ramos, and Bratt}]{franco1999automatic}
\bibinfo{author}{Franco, H.}, \bibinfo{author}{Neumeyer, L.}, \bibinfo{author}{Ramos, M.},  and \bibinfo{author}{Bratt, H.} (\textbf{\bibinfo{year}{1999}}). \enquote{\bibinfo{title}{Automatic detection of phone-level mispronunciation for language learning}} \bibinfo{journal}{Conf. on Speech Communication and Technology,} \bibinfo{pages}{pp. 851--854}, \dodoi{10.21437/Eurospeech.1999-207}.

\bibitem[{Ghesquiere and Van~Compernolle(2002)}]{ghesquiere2002flemish}
\bibinfo{author}{Ghesquiere, P.-J.},  and \bibinfo{author}{Van~Compernolle, D.} (\textbf{\bibinfo{year}{2002}}). \enquote{\bibinfo{title}{Flemish accent identification based on formant and duration features}} \textbf{1}, \bibinfo{pages}{pp. 749--752}, \dodoi{10.1109/icassp.2002.1005848}.

\bibitem[{Ghorbani and Hansen(2018)}]{cp1}
\bibinfo{author}{Ghorbani, S.},  and \bibinfo{author}{Hansen, J. H.~L.} (\textbf{\bibinfo{year}{2018}}). \enquote{\bibinfo{title}{Leveraging native language information for improved accented speech recognition}} \bibinfo{journal}{ISCA Interspeech,} \bibinfo{pages}{pp. 2449--2453}, \dodoi{10.21437/Interspeech.2018-1378}.

\bibitem[{Ghorbani and Hansen(2023)}]{domainexpansion}
\bibinfo{author}{Ghorbani, S.},  and \bibinfo{author}{Hansen, J. H.~L.} (\textbf{\bibinfo{year}{2023}}). \enquote{\bibinfo{title}{Domain expansion for end-to-end speech recognition: Applications for accent/dialect speech}} \bibinfo{journal}{IEEE Transactions on Audio, Speech, and Language Processing} \textbf{31}, \bibinfo{pages}{pp. 762--774}, \dodoi{10.1109/TASLP.2022.3233238}.

\bibitem[{Graves \emph{et~al.}(2006)Graves, Fern{\'a}ndez, Gomez, and Schmidhuber}]{CTC}
\bibinfo{author}{Graves, A.}, \bibinfo{author}{Fern{\'a}ndez, S.}, \bibinfo{author}{Gomez, F.},  and \bibinfo{author}{Schmidhuber, J.} (\textbf{\bibinfo{year}{2006}}). \enquote{\bibinfo{title}{Connectionist temporal classification: labelling unsegmented sequence data with recurrent neural networks}} \bibinfo{journal}{Inter. Conf. on Machine Learning (ICML),} \bibinfo{pages}{pp. 369--376}, \dodoi{10.1145/1143844.1143891}.

\bibitem[{Gulati \emph{et~al.}(2020)Gulati, Qin, Chiu, Parmar, Zhang, Yu, Han, Wang, Zhang, Wu \emph{et~al.}}]{conformer}
\bibinfo{author}{Gulati, A.}, \bibinfo{author}{Qin, J.}, \bibinfo{author}{Chiu, C.-C.}, \bibinfo{author}{Parmar, N.}, \bibinfo{author}{Zhang, Y.}, \bibinfo{author}{Yu, J.}, \bibinfo{author}{Han, W.}, \bibinfo{author}{Wang, S.}, \bibinfo{author}{Zhang, Z.}, \bibinfo{author}{Wu, Y.} \emph{et~al.} (\textbf{\bibinfo{year}{2020}}). \enquote{\bibinfo{title}{Conformer: Convolution-augmented transformer for speech recognition}} \bibinfo{journal}{ISCA Interspeech,} \bibinfo{pages}{pp. 5036--5040}, \dodoi{10.21437/interspeech.2020-3015}.

\bibitem[{Huang \emph{et~al.}(2021)Huang, Xiang, Yang, Ma, and Qian}]{huang2021aispeech}
\bibinfo{author}{Huang, H.}, \bibinfo{author}{Xiang, X.}, \bibinfo{author}{Yang, Y.}, \bibinfo{author}{Ma, R.},  and \bibinfo{author}{Qian, Y.} (\textbf{\bibinfo{year}{2021}}). \enquote{\bibinfo{title}{Aispeech-{SJTU} accent identification system for the accented {E}nglish speech recognition challenge}} \bibinfo{journal}{IEEE Inter. Conf. Acoustics, Speech, and Signal Proc. (ICASSP),} \bibinfo{pages}{pp. 6254--6258}, \dodoi{10.1109/icassp39728.2021.9414292}.

\bibitem[{Jiao \emph{et~al.}(2016)Jiao, Tu, Berisha, and Liss}]{jiao2016accent}
\bibinfo{author}{Jiao, Y.}, \bibinfo{author}{Tu, M.}, \bibinfo{author}{Berisha, V.},  and \bibinfo{author}{Liss, J.~M.} (\textbf{\bibinfo{year}{2016}}). \enquote{\bibinfo{title}{Accent identification by combining deep neural networks and recurrent neural networks trained on long and short term features}} \bibinfo{journal}{ISCA Interspeech,} \bibinfo{pages}{pp. 2388--2392}, \dodoi{10.21437/interspeech.2016-1148}.

\bibitem[{Kang(2010)}]{kang2010relative}
\bibinfo{author}{Kang, O.} (\textbf{\bibinfo{year}{2010}}). \enquote{\bibinfo{title}{Relative salience of suprasegmental features on judgments of {L2} comprehensibility and accentedness}} \bibinfo{journal}{System} \textbf{38}(2), \bibinfo{pages}{pp. 301--315}, \dodoi{10.1016/j.system.2010.01.005}.

\bibitem[{Kang \emph{et~al.}(2018)Kang, Thomson, and Moran}]{kang2020}
\bibinfo{author}{Kang, O.}, \bibinfo{author}{Thomson, R.~I.},  and \bibinfo{author}{Moran, M.} (\textbf{\bibinfo{year}{2018}}). \enquote{\bibinfo{title}{Which features of accent affect understanding? exploring the intelligibility threshold of diverse accent varieties}} \bibinfo{journal}{Applied Linguistics} \textbf{41}(4), \bibinfo{pages}{pp. 453--480}, \dodoi{10.1093/applin/amy053}.

\bibitem[{Kennedy and Trofimovich(2008)}]{kennedy2008}
\bibinfo{author}{Kennedy, S.},  and \bibinfo{author}{Trofimovich, P.} (\textbf{\bibinfo{year}{2008}}). \enquote{\bibinfo{title}{Intelligibility, comprehensibility, and accentedness of {L2} speech: The role of listener experience and semantic context}} \bibinfo{journal}{Canadian Modern Language Review} \textbf{64}(3), \bibinfo{pages}{pp. 459--489}, \dodoi{10.3138/cmlr.64.3.459}.

\bibitem[{Kingma and Ba(2015)}]{kingma2014adam}
\bibinfo{author}{Kingma, D.~P.},  and \bibinfo{author}{Ba, J.} (\textbf{\bibinfo{year}{2015}}). \enquote{\bibinfo{title}{Adam: {A} method for stochastic optimization}} \bibinfo{journal}{Inter. Conf. on Learning Representations(ICLR),} \bibinfo{pages}{pp. 1--13}.

\bibitem[{Kukk and Alum{\"a}e()}]{LIDaccent}
\bibinfo{author}{Kukk, K.},  and \bibinfo{author}{Alum{\"a}e, T.} \enquote{\bibinfo{title}{Improving language identification of accented speech}} \bibinfo{journal}{ISCA Interspeech,} \bibinfo{pages}{pp. 1288--1292}, \dodoi{10.21437/Interspeech.2022-10455}.

\bibitem[{Nagrani \emph{et~al.}(2017)Nagrani, Chung, and Zisserman}]{voxceleb1}
\bibinfo{author}{Nagrani, A.}, \bibinfo{author}{Chung, J.~S.},  and \bibinfo{author}{Zisserman, A.} (\textbf{\bibinfo{year}{2017}}). \enquote{\bibinfo{title}{{VoxCeleb}: A large-scale speaker identification dataset}} \bibinfo{journal}{ISCA Interspeech,} \bibinfo{pages}{pp. 2616--2620}, \dodoi{10.21437/Interspeech.2017-950}.

\bibitem[{Nam \emph{et~al.}(2022)Nam, Kim, Heo, Jung, and Son~Chung}]{disentangled}
\bibinfo{author}{Nam, K.}, \bibinfo{author}{Kim, Y.}, \bibinfo{author}{Heo, H.~S.}, \bibinfo{author}{Jung, J.-w.},  and \bibinfo{author}{Son~Chung, J.} (\textbf{\bibinfo{year}{2022}}). \enquote{\bibinfo{title}{Disentangled representation learning for multilingual speaker recognition}} \bibinfo{journal}{arXiv:1412.6980. [Online]. Available: https://arxiv.org/abs/1412.6980} .

\bibitem[{Panayotov \emph{et~al.}(2015)Panayotov, Chen, Povey, and Khudanpur}]{librispeech}
\bibinfo{author}{Panayotov, V.}, \bibinfo{author}{Chen, G.}, \bibinfo{author}{Povey, D.},  and \bibinfo{author}{Khudanpur, S.} (\textbf{\bibinfo{year}{2015}}). \enquote{\bibinfo{title}{{L}ibrispeech: an {ASR} corpus based on public domain audio books}} \bibinfo{journal}{IEEE Inter. Conf. Acoustics, Speech, and Signal Proc. (ICASSP),} \bibinfo{pages}{pp. 5206--5210}, \dodoi{10.1109/icassp.2015.7178964}.

\bibitem[{Pedregosa \emph{et~al.}(2011)Pedregosa, Varoquaux, Gramfort, Michel, Thirion, Grisel, Blondel, Prettenhofer, Weiss, Dubourg, Vanderplas, Passos, Cournapeau, Brucher, Perrot, and Duchesnay}]{scikit-learn}
\bibinfo{author}{Pedregosa, F.}, \bibinfo{author}{Varoquaux, G.}, \bibinfo{author}{Gramfort, A.}, \bibinfo{author}{Michel, V.}, \bibinfo{author}{Thirion, B.}, \bibinfo{author}{Grisel, O.}, \bibinfo{author}{Blondel, M.}, \bibinfo{author}{Prettenhofer, P.}, \bibinfo{author}{Weiss, R.}, \bibinfo{author}{Dubourg, V.}, \bibinfo{author}{Vanderplas, J.}, \bibinfo{author}{Passos, A.}, \bibinfo{author}{Cournapeau, D.}, \bibinfo{author}{Brucher, M.}, \bibinfo{author}{Perrot, M.},  and \bibinfo{author}{Duchesnay, E.} (\textbf{\bibinfo{year}{2011}}). \enquote{\bibinfo{title}{Scikit-learn: Machine learning in {P}ython}} \bibinfo{journal}{Journal of Machine Learning Research} \textbf{12}, \bibinfo{pages}{2825--2830}.

\bibitem[{Piat \emph{et~al.}(2008)Piat, Fohr, and Illina}]{piat2008foreign}
\bibinfo{author}{Piat, M.}, \bibinfo{author}{Fohr, D.},  and \bibinfo{author}{Illina, I.} (\textbf{\bibinfo{year}{2008}}). \enquote{\bibinfo{title}{Foreign accent identification based on prosodic parameters}} \bibinfo{journal}{ISCA Interspeech,} \bibinfo{pages}{pp. 759--762}, \dodoi{10.21437/Interspeech.2008-235}.

\bibitem[{Porretta and Tucker(2012)}]{porretta2012predicting}
\bibinfo{author}{Porretta, V.},  and \bibinfo{author}{Tucker, B.~V.} (\textbf{\bibinfo{year}{2012}}). \enquote{\bibinfo{title}{Predicting accentedness: Acoustic measurements of {C}hinese-accented {E}nglish}} \bibinfo{journal}{Canadian Acoustics} \textbf{40}(3), \bibinfo{pages}{pp. 34--35}.

\bibitem[{Raj \emph{et~al.}(2019)Raj, Snyder, Povey, and Khudanpur}]{probing}
\bibinfo{author}{Raj, D.}, \bibinfo{author}{Snyder, D.}, \bibinfo{author}{Povey, D.},  and \bibinfo{author}{Khudanpur, S.} (\textbf{\bibinfo{year}{2019}}). \enquote{\bibinfo{title}{Probing the information encoded in {x}-vectors}} \bibinfo{journal}{IEEE Automatic Speech Recog. and Understanding Workshop (ASRU),} \bibinfo{pages}{pp. 726--733}, \dodoi{10.1109/asru46091.2019.9003979}.

\bibitem[{Ranjan and Hansen(2018)}]{CL-SID}
\bibinfo{author}{Ranjan, S.},  and \bibinfo{author}{Hansen, J. H.~L.} (\textbf{\bibinfo{year}{2018}}). \enquote{\bibinfo{title}{Curriculum {L}earning based approaches for noise robust speaker recognition}} \bibinfo{journal}{IEEE Transactions on Audio, Speech, and Language Processing} \textbf{26}(1), \bibinfo{pages}{pp. 197--210}, \dodoi{10.1109/TASLP.2017.2765832}.

\bibitem[{Ravanelli \emph{et~al.}(2021)Ravanelli, Parcollet, Plantinga, Rouhe, Cornell, Lugosch, Subakan, Dawalatabad, Heba, Zhong, Chou, Yeh, Fu, Liao, Rastorgueva, Grondin, Aris, Na, Gao, Mori, and Bengio}]{speechbrain}
\bibinfo{author}{Ravanelli, M.}, \bibinfo{author}{Parcollet, T.}, \bibinfo{author}{Plantinga, P.}, \bibinfo{author}{Rouhe, A.}, \bibinfo{author}{Cornell, S.}, \bibinfo{author}{Lugosch, L.}, \bibinfo{author}{Subakan, C.}, \bibinfo{author}{Dawalatabad, N.}, \bibinfo{author}{Heba, A.}, \bibinfo{author}{Zhong, J.}, \bibinfo{author}{Chou, J.-C.}, \bibinfo{author}{Yeh, S.-L.}, \bibinfo{author}{Fu, S.-W.}, \bibinfo{author}{Liao, C.-F.}, \bibinfo{author}{Rastorgueva, E.}, \bibinfo{author}{Grondin, F.}, \bibinfo{author}{Aris, W.}, \bibinfo{author}{Na, H.}, \bibinfo{author}{Gao, Y.}, \bibinfo{author}{Mori, R.~D.},  and \bibinfo{author}{Bengio, Y.} (\textbf{\bibinfo{year}{2021}}). \enquote{\bibinfo{title}{{SpeechBrain}: A general-purpose speech toolkit}} \bibinfo{journal}{arXiv:2106.04624 [Online]. Available: https://arxiv.org/abs/2106.04624} \bibinfo{note}{ArXiv:2106.04624}.

\bibitem[{Smith(2017)}]{cyclical}
\bibinfo{author}{Smith, L.~N.} (\textbf{\bibinfo{year}{2017}}). \enquote{\bibinfo{title}{Cyclical learning rates for training neural networks}} \bibinfo{journal}{IEEE Winter Conf. on Applications of Computer Vision (WACV),} \bibinfo{pages}{pp. 464--472}, \dodoi{10.1109/wacv.2017.58}.

\bibitem[{Snyder \emph{et~al.}(2018)Snyder, Garcia-Romero, Sell, Povey, and Khudanpur}]{x-vector}
\bibinfo{author}{Snyder, D.}, \bibinfo{author}{Garcia-Romero, D.}, \bibinfo{author}{Sell, G.}, \bibinfo{author}{Povey, D.},  and \bibinfo{author}{Khudanpur, S.} (\textbf{\bibinfo{year}{2018}}). \enquote{\bibinfo{title}{X-vectors: Robust dnn embeddings for speaker recognition}} \bibinfo{journal}{IEEE Inter. Conf. Acoustics, Speech, and Signal Proc. (ICASSP),} \bibinfo{pages}{pp. 5329--5333}, \dodoi{10.1109/icassp.2018.8461375}.

\bibitem[{Valk and Alum{\"a}e(2021)}]{valk2021slt}
\bibinfo{author}{Valk, J.},  and \bibinfo{author}{Alum{\"a}e, T.} (\textbf{\bibinfo{year}{2021}}). \enquote{\bibinfo{title}{{VoxLingua107}: a dataset for spoken language recognition}} \bibinfo{journal}{IEEE SLT Workshop,} \bibinfo{pages}{pp. 652--658}, \dodoi{10.1109/slt48900.2021.9383459}.

\bibitem[{van Maastricht \emph{et~al.}(2016)van Maastricht, Krahmer, and Swerts}]{van2016native}
\bibinfo{author}{van Maastricht, L.}, \bibinfo{author}{Krahmer, E.},  and \bibinfo{author}{Swerts, M.} (\textbf{\bibinfo{year}{2016}}). \enquote{\bibinfo{title}{Native speaker perceptions of (non-) native prominence patterns: Effects of deviance in pitch accent distributions on accentedness, comprehensibility, intelligibility, and nativeness}} \bibinfo{journal}{Speech Communication} \textbf{83}, \bibinfo{pages}{pp. 21--33}, \dodoi{10.1016/j.specom.2016.07.008}.

\bibitem[{Viglino \emph{et~al.}(2019)Viglino, Motlicek, and Cernak}]{viglino2019end}
\bibinfo{author}{Viglino, T.}, \bibinfo{author}{Motlicek, P.},  and \bibinfo{author}{Cernak, M.} (\textbf{\bibinfo{year}{2019}}). \enquote{\bibinfo{title}{End-to-end accented speech recognition}} \bibinfo{journal}{ISCA Interspeech} \bibinfo{pages}{2140--2144}.

\bibitem[{Waskom(2021)}]{seaborn}
\bibinfo{author}{Waskom, M.~L.} (\textbf{\bibinfo{year}{2021}}). \enquote{\bibinfo{title}{seaborn: statistical data visualization}} \bibinfo{journal}{Journal of Open Source Software} \textbf{6}(60), \bibinfo{pages}{3021}, \dourl{https://doi.org/10.21105/joss.03021}, \dodoi{10.21105/joss.03021}.

\bibitem[{Weninger \emph{et~al.}(2019)Weninger, Sun, Park, Willett, and Zhan}]{weninger2019deep}
\bibinfo{author}{Weninger, F.}, \bibinfo{author}{Sun, Y.}, \bibinfo{author}{Park, J.}, \bibinfo{author}{Willett, D.},  and \bibinfo{author}{Zhan, P.} (\textbf{\bibinfo{year}{2019}}). \enquote{\bibinfo{title}{Deep learning based mandarin accent identification for accent robust {ASR}}} \bibinfo{journal}{ISCA Interspeech,} \bibinfo{pages}{pp. 510--514}, \dodoi{10.21437/interspeech.2019-2737}.

\bibitem[{Witt and Young(2000)}]{witt2000phone}
\bibinfo{author}{Witt, S.~M.},  and \bibinfo{author}{Young, S.~J.} (\textbf{\bibinfo{year}{2000}}). \enquote{\bibinfo{title}{Phone-level pronunciation scoring and assessment for interactive language learning}} \bibinfo{journal}{Speech communication} \textbf{30}(2-3), \bibinfo{pages}{pp. 95--108}, \dodoi{10.1016/s0167-6393(99)00044-8}.

\bibitem[{Wu \emph{et~al.}(2021)Wu, Li, Leung, and Meng}]{wu2021transformer}
\bibinfo{author}{Wu, M.}, \bibinfo{author}{Li, K.}, \bibinfo{author}{Leung, W.-K.},  and \bibinfo{author}{Meng, H.} (\textbf{\bibinfo{year}{2021}}). \enquote{\bibinfo{title}{Transformer based end-to-end mispronunciation detection and diagnosis}} \bibinfo{journal}{ISCA Interspeech,} \bibinfo{pages}{pp. 3954--3958}, \dodoi{10.21437/Interspeech.2021-1467}.

\bibitem[{Yang \emph{et~al.}(2022)Yang, Hirschi, Looney, Kang, and Hansen}]{yang2022improving}
\bibinfo{author}{Yang, M.}, \bibinfo{author}{Hirschi, K.}, \bibinfo{author}{Looney, S.~D.}, \bibinfo{author}{Kang, O.},  and \bibinfo{author}{Hansen, J. H.~L.} (\textbf{\bibinfo{year}{2022}}). \enquote{\bibinfo{title}{Improving mispronunciation detection with wav2vec2-based momentum pseudo-labeling for accentedness and intelligibility assessment}} \bibinfo{journal}{ISCA Interspeech,} \bibinfo{pages}{pp. 4481--4485}, \dodoi{10.21437/interspeech.2022-11039}.

\end{thebibliography}

\end{document}